\documentclass[noshowpacs,onecolumn,showpacs,superscriptaddress,groupedaddress,amsmath,amssymb]{elsarticle}

\usepackage{color}
\usepackage{amsmath,amssymb}
\usepackage{pifont}
\usepackage{amssymb}  
\usepackage{bbold}
\usepackage{float}
\usepackage{subfig}
\usepackage{tikz}
\usepackage[left=2cm,right=2cm,bottom=2cm,top=2cm]{geometry}
\usepackage{makecell}
\usepackage{pifont}   
\usepackage{graphicx} 
\usepackage{dcolumn}  
\usepackage{bm}       
\usepackage{multirow} 
\usepackage[colorlinks]{hyperref}
\usepackage{placeins}
\newcommand{\vect}[1]{\mathbf{#1}}
\newcommand{\ket}[1]{\left|{#1}\right\rangle}

\newcommand{\eqfitpage}[1]{\resizebox{\linewidth}{!}{$#1$}}

\newcommand{\hiddensubsubsection}[1]{
    \stepcounter{subsubsection}
    \subsection*{\arabic{section}.\arabic{subsection}.\arabic{subsubsection}\hspace{1em}{#1}}
}

\biboptions{sort&compress}

\begin{document}
\title{Tunneling in an anisotropic cubic Dirac semi-metal}
\author[1]{Ahmed Bouhlal}
\author[1]{Ahmed Jellal}
\author[2]{Hocine Bahlouli}
\author[2]{Michael Vogl}
\ead{ssss133@googlemail.com}
\address[1]{Laboratory of Theoretical Physics, Faculty of Sciences, Choua\"ib Doukkali University,  24000 El Jadida, Morocco}
\address[2]{Department of Physics, King Fahd University of Petroleum and Minerals, 31261 Dhahran, Saudi Arabia}
\begin{abstract}
Motivated by a recent first principles prediction of an anisotropic cubic Dirac semi-metal in a real material Tl(TeMo)$_3$, we study the behavior of electrons tunneling through a potential barrier in such systems. To clearly investigate effects from different contributions to the Hamiltonian we study the model in various limits. First, in the limit of a very thin film material where the linearly dispersive $z$-direction is frozen out at zero momentum and the dispersion in the $x$-$y$ plane is rotationally symmetric. In this limit we find a Klein tunneling reminiscent of what is observed in single layer graphene and linearly dispersive Dirac semi-metals. Second, an increase in thickness of the material leads to the possibility of a non-zero momentum eigenvalue $k_z$ that acts as an effective mass term in the Hamiltonian. We find that these lead to a suppression of Klein tunneling. Third, the inclusion of an anisotropy parameter $\lambda\neq 1$ leads to a breaking of rotational invariance. Furthermore, we observed that for different values of incident angle $\theta$ and anisotropy parameter $\lambda$ the Hamiltonian supports different numbers of modes propagating to infinity. We display this effect in form of a diagram that is similar to a phase diagram of a distant detector. Fourth, we consider coexistence of both anisotropy and non-zero $k_z$ but do not find any  effect that is unique to the interplay between non-zero momentum $k_z$ and anisotropy parameter $\lambda$. Last, we studied the case of a barrier that was placed in the linearly dispersive direction and found Klein tunneling $T-1\propto \theta^6+\mathcal{O}(\theta^8)$ that is enhanced when compared to the Klein tunneling in linear Dirac semi-metals or graphene where $T-1\propto \theta^2+\mathcal{O}(\theta^4)$. 
\end{abstract}
\date{\today}
\maketitle
\tableofcontents

\section{Introduction}
After the discovery of graphene \cite{novoselov2004electric} - a two-dimensional lattice of carbon atoms that are arranged in a honeycomb pattern forming a hexagonal lattice - two dimensional materials have become a hot topic both experimentally and theoretically. After all, graphene hosts overwhelmingly many exotic and intriguing phenomena \cite{Castro_Neto_2009,Geim_2009,zhang2005experimental, chung2010transferable,Katsnelson_2006,vogl2017semiclassics,katsnelson2012graphene,logemann2015modeling,young2012spin,klier2015ruderman,bistritzer2009electronic,nomura2006quantum,walter2011effective,xiao2007valley,carmier2008berry,rodriguez2014ground,lu2015local,rodriguez2017giant,kormanyos2008bound,reijnders2018electronic,vogl2020effective_TBG,perez2014floquet,usaj2014irradiated,calvo2011tuning,torres2020introduction,pereira2007graphene,dayi2010noncommutative,jellal2021measuring,mekkaoui2020fano,abdullah2018graphene,abdullah2018confined,berdiyorov2016effect,zahidi2020magnetic,partoens2006graphene}. This and the many promising and fascinating physical, electrical, chemical, magnetic and optical properties\cite{Giustino_2021,schaibley2016valleytronics,novoselov20162d,avouris20172d,gibertini2019magnetic,mas20112d,ryu2019superlattices,wallace2014situ,choudhuri2019recent} have since  motivated considerable efforts that were deployed to study properties of other novel 2D materials. Examples include similar thin materials such as silicene\cite{vogt2012silicene,ni2012tunable,kara2012review,kara2009physics,lalmi2010epitaxial}, transition metal dichalcogenides\cite{Xiao_2012,PhysRevB.88.085433,manzeli20172d,wang2012electronics,liu2014optical,friend1987electronic,voiry2015phase,chhowalla2015two,azhikodan2016anomalous} and very importantly a large amount of work on multi-layer materials\cite{min2008electronic,Wu_2018,Bistritzer12233,Cao2018sc,Codecidoeaaw9770,Yankowitz1059,chichinadze2019nematic,fleischmann2019moir,Wu_2019,zhai2020theory,zhang2020tuning,zhan2020multiultraflatbands,venkateswarlu2020electronic,chen2020configure,San_Jose_2012,zhang2020correlated,xie2020topology,wu2020collective,shallcross2008quantum,vogl2020floquet,rost2019nonperturbative,jung2011valley,min2008chiral,zhu2014optical,abdullah2018electronic,abdullah2016gate,zhu2020twisted,ma2020topological,shi2020tunable,zuo2018scanning,amorim2018electronic,lopez2020electrical,lei2020mirror,li2019electronic,mora2019flatbands,vitale2021flat} to name a few. 

However, it was soon realized that the peculiar properties of graphene are not just a consequence of its low dimensionality but rather due to its linear band structure. This naturally led to a search for similar materials in higher dimensions \cite{Armitage_2018, doi:10.1146/annurev-conmatphys-031016-025458} and other generalizations that all are now characterized as Dirac and Weyl semi-metals.

Weyl semi-metals \cite{doi:10.1146/annurev-conmatphys-031016-025458, wang2017quantum, PhysRevX.7.021019,zyuzin2016intrinsic,jiang2017signature,wang2016mote,weng2015weyl,xu2015discovery,soluyanov2015type,huang2015weyl,xu2015discovery,weng2015weyl,huang2015observation,zyuzin2012weyl,lundgren2014thermoelectric,lundgren2015electronic,bulmash2014prediction,jiang2012tunable,burkov2011weyl,wang2020higher,ghorashi2020higher,wang2014floquet,zhang2016theory,hubener2017creating,chen2016thermoelectric,lv2015experimental,liu2019magnetic,vazifeh2013electromagnetic,Alidoust_2020,Alidoust_2018,Halterman_2019,Halterman_2018,Alidoust_2017,huang2016new} and Dirac semi-metals \cite{Armitage_2018, doi:10.1146/annurev-conmatphys-031016-025458, wang2017quantum,lundgren2014thermoelectric,lundgren2015electronic,hubener2017creating,wan2011topological,xu2011chern,fang2012multi,xiong2015evidence,liu2014stable,wang2013three,liang2015ultrahigh,novak2015large,wieder2016double,he2014quantum,jeon2014landau,weeks2010interaction,kim2015observation,li2018giant,chang2017type,szabo2020dirty,wieder2020strong,wang2012dirac,young2012dirac,liu2014discovery,yang2014classification,young2015dirac,borisenko2014experimental} are considered to be the 3D analogues of graphene materials. The common feature that is shared by linear 3D Dirac and Weyl semi-metals is the fact that conduction and valence bands touch each other at single, discrete points in $k$-space, around which the dispersion has a linear $k$-dependence. This unusual band dispersion causes the electrons around the Fermi energy to behave like relativistic particles similarly to graphene systems. However, band touching points need not necessarily be linear and higher order band-crossings are possible \cite{PhysRevX.7.021019,fang2012multi,huang2016new,chen2016thermoelectric,yang2014classification}. Even more generally, band crossings in Dirac and Weyl semimetals  (called “Dirac" or "Weyl" points) can even have different order of dispersion in different directions as dictated by their crystal symmetries. It has been realized recently that these Dirac and Weyl band-crossing points carry important information on the topological behavior of the system, giving rise to specific behaviors of surface or edge states in such systems \cite{ 4c72726ab83a4fdb83509d8bb9f6c45c,novoselov2004electric, PhysRevX.7.021019}. These gapless degrees of freedom are protected by the topological properties of the system and define the so-called topological invariants of the system. Thus Dirac and Weyl semimetals possess non-trivial protected topological properties that are at the origin of some of their exciting electronic and transport properties. Of course there exists a large zoo of similar materials that might not just have band touching points but where bands might even touch along nodal lines or on surfaces \cite{Fang_2016,xiao2020experimental,qie2019tetragonal,zhang2018nodal,xiao2017topologically,burkov2011topological,meng2020nonsymmorphic,wang2018pseudo,emmanouilidou2017magnetotransport,li2019new,huang2016topological,li2018evidence,yu2017topological,he2018type,li2018orthorhombic,laha2020magnetotransport,wu2018nodal,zhang2017topological,hosen2017tunability,chang2019realization}. Many semi-metals have been studied in some detail including their tunneling properties but not all of them. Particularly, not all semi-metals with higher order band crossing points have been studied exhaustively. Some recent examples that have been studied include multi Weyl semi-metals with a dispersion that is rotationally symmetric in the x-y plane \cite{mandal2021tunneling,Mandal_2020,bera2021floquet} and higher psuedo-spin generalizations of Weyl semi-metals \cite{Mandal_2020higher_spin}.

In our present work we study the behavior of electrons tunneling through a potential barrier in an “anisotropic cubic Dirac fermion” system like the one that was recently predicted based on first principles computations for Tl(TeMo)$_3$ and related materials, which are real and stable solid state systems \cite{PhysRevX.7.021019}. It should be stressed that we study a dispersion that is also anisotropic in the  x-y plane but keeps a sixfold rotational symmetry.

The work is organized as follows. In section \ref{sec:Model} we introduce the mathematical model we consider, in section \ref{sec:2dRotSymm_limit} we study the 2D limit of a thin material that has momentum in $z$-direction frozen out at zero value (setting $k_z=0$) and therefore behaves approximately as a two-dimensional system. Here we find Klein tunneling behaviour that is very reminiscent of what is known in graphene or linearly dispersive semi metals. In section \ref{sec:kzNotZeroIsotropic}, we consider the case of a cubic Dirac semi metal that is isotropic in the plane but allows a finite momentum $k_z$
in $z$-direction. The momentum term acts similarly to a mass term and therefore spoils some of the Klein tunneling properties. In section \ref{Sec:anisotropkz0} we consider the anisotropic case with $k_z=0$, where we discover that for different angles and anisotropy strengths tunneling properties can differ wildly. In section \ref{sec:anisotrop_kz_not_zero} we consider the anisotropic case but with a non zero $k_z$ value and the addition of the mass-term mainly serves to damp tunneling phenomena. Lastly, in section \ref{sec:tunneling_z_dir} 
we consider tunneling through a potential barrier along the linearly dispersive $z$-direction in a bulk cubic Dirac semi-metal, where we discover Klein tunneling that is enhanced compared to the Klein tunneling that is known from graphene and linearly dispersive semi-metals. Finally, in \ref{sec:conclusion} we present our conclusion concerning our main findings and discuss some open questions.

\section{Theoretical model}
\label{sec:Model}

The Hamiltonian describing an anisotropic cubic Dirac semi-metal can be written as follows \cite{PhysRevX.7.021019} 
\begin{equation}
  \label{eq:Hamiltonian}
  H(k) = \hbar \left(\begin{array}{cc} h(k)&0\\0&-h(k)\\\end{array}\right)+V(x) \mathbb{I}_{4},
\end{equation}
where
\begin{equation}
\label{eq:unite}
h(k)=v_x\sigma_x\left(\hat{k}_{+}^{3}+\hat{k}_{-}^{3}\right)+iv_y\sigma_y\left(\hat{k}_{+}^{3}-\hat{k}_{-}^{3}\right)+v_zk_z\sigma_z
\end{equation}
An isotropic version of this model was also presented in \cite{yang2014classification} and the terms also appears in the context of spin-orbit coupling corrections \cite{Alidoust_2021}. In the equation, we have $\hat{k}_i$ as momentum operators. We used the shorthand notation $\hat{k}_{\pm}=\hat k_x \pm i \hat k_y$, $\sigma_i$ are Pauli matrices and $v_{x,y,z}$ are independent real coefficients with dimension $T^{-1}$ (T stands for time). That is,  wavevector $\vect k$ and positions $\vect x$ are chosen dimensionless. Typically in the real solid state system that realizes the above Hamiltonian it is written in a basis set that corresponds to a wavefunction of the form $\ket{\psi}=(\ket{A,\uparrow},\ket{B,\uparrow},\ket{A,\downarrow},\ket{B,\downarrow})^T$, where $A$ and $B$ denote some orbital degrees of freedom and $\uparrow/\downarrow$ spin degrees of freedom \cite{PhysRevX.7.021019}.

We consider the case where the system is subject to a scalar potential barrier of the form
\begin{equation}\label{eV0}
	V(x)=
	\left\{%
	\begin{array}{ll}
		 V_0, & -\frac{L}{2}<x<\frac{L}{2}\\
         0,  & \mbox{elsewhere} \\
	\end{array}%
	\right.,
\end{equation}

which is also shown in Fig. \ref{fig:barrier_scheme}:
 \begin{figure}[H]
 \centering
\includegraphics[width=0.5\linewidth]{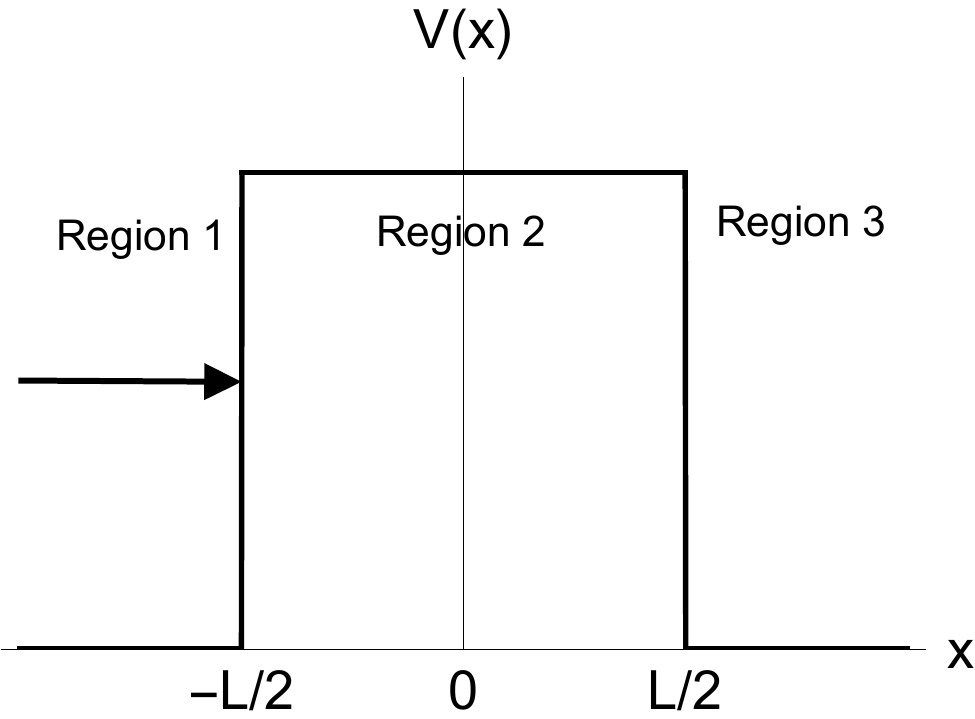}
\caption{The figure schematically shows the setting of our scattering problem and introduces the labeling scheme for the different potential regions. }
\label{fig:barrier_scheme}
\end{figure}

This type of Hamiltonian has several interesting properties, which are the cubic dispersion, anisotropy in the $x$-$y$ plane and linear momentum dependence in $z$-direction. To fully understand the effect of all these contributions to the tunneling phenomena we will study the effects of each of these contributions separately. 
That is, we will first study simplest case of the rotationally symmetric 2D limit $v_x=v_y$ with $k_z=0$. Afterwards we will investigate the effects of non-zero $k_z$, then the anisotropic limit $v_x\neq v_y$ with $k_z=0$ and finally the case of both $k_z\neq 0$ and $v_x\neq v_y$.


\section{Analysis of different limits}
\subsection{Isotropic 2D \texorpdfstring{$(k_z=0)$}{(kz=0)} limit}
\label{sec:2dRotSymm_limit}
As a first step and to gain some intuition about the problem at hand, we consider the $2D$ flat limit by setting $k_z=0$. Furthermore, we consider the special case of an isotropic dispersion setting $v_x=v_y=v$ for simplicity. The limit of $k_z=0$ is relevant because typically $v_z\gg v_{x,y}$ and therefore $k_z$ is typically frozen out if the sample is sufficiently thin (see appendix \ref{app:freeze_out_kz} for details).

Under these simplifying conditions one may write the Hamiltonian of the system in a matrix form 
 \begin{equation}
  H(k) = \left(\begin{array}{cccc} V(x)&2\hbar v \hat{k}_{+}^3&0&0\\2\hbar v \hat{k}_{-}^3&V(x)&0&0\\0&0&V(x)&-2\hbar v \hat{k}_{+}^3\\0&0&-2\hbar v \hat{k}_{-}^3&V(x)\\\end{array}\right).
\end{equation}
In order to solve this eigenvalue problem we can separate variables and write the eigenspinors as
plane waves in the $y$-direction. This is due to the fact that $[H,k_y]=0$ implies conservation
of momentum along the $y$-direction $\Psi(x,y)= e^{ik_y y} \psi(x)$. 

\begin{figure}[H]
  \centering
\includegraphics[width=0.5\linewidth]{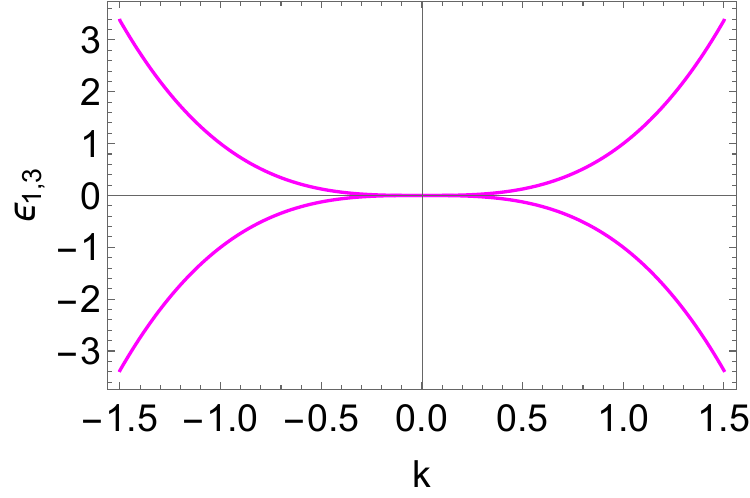}~~\includegraphics[width=0.5\linewidth]{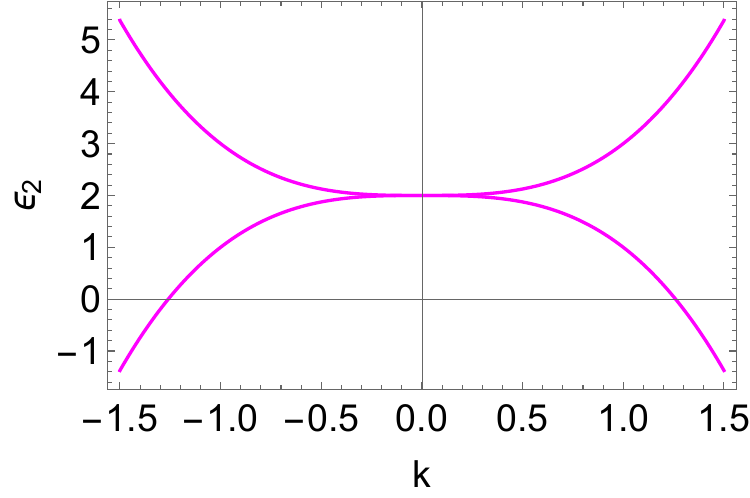}
\caption{\sf (color online) Reduced energy $\epsilon$ in different regions as a function of the incident wavevector amplitude $k$. The left panel shows regions 1 and 3 and the right panel shows region 2. } 
\label{fig:dispersion_symmetric_delta0u2}
\end{figure}

Solutions in each region are then given by plane waves $\psi(x)\propto e^{\pm i k_x x}$.
It is convenient to switch to dimensionless variables by defining the reduced energy $\epsilon$ and potential $u$
\begin{equation}
\epsilon=\frac{E}{2 \hbar v};\quad u_j=\frac{V_j}{2 \hbar v}
\label{eq:unitless_en}
\end{equation}
where $V_j$ is the potential in region $j$ (compare Fig. \ref{fig:barrier_scheme}).
The reduced energy eigenvalues in each region therefore are given by  
\begin{equation}
\epsilon_j=\pm(k_x^2+k_y^2)^\frac{3}{2}+u_j.
\end{equation}
Since this reduced energy formula is rotationally symmetric the result can be plotted as a function of the wavevector amplitude $k=(k_x^2+k_y^2)^{\frac{1}{2}}$, which is shown in the Fig. \ref{fig:dispersion_symmetric_delta0u2}.
We find that the band touching point is very flat which means that kinetic energy of carriers is very small in these regions and hence interactions dominate these regions.

\hiddensubsubsection{Solution of the tunneling problem}
\label{sec:Sol:wf:massless_isotrol}
We now turn to the tunneling problem at a fixed value of $\epsilon$. The energy $\epsilon$ is determined by the wavevector $\vect k=(\cos\theta,\sin\theta)k$ of the incoming wave. It is therefore convenient to recast the problem as a \textit{generalized cubic eigenvalue problem} for $k_x$ rather than a linear energy eigenvalue problem for $\epsilon$. This is because the problem is more clearly understood in this language. 

After all, for the tunneling problem at hand one needs to match wavefunctions, their derivatives and second derivatives to ensure smoothness because the Hamiltonian includes up to third order derivatives. At each boundary region there are four matching conditions (one for each vector component). The same is true for the matching conditions of first and second derivatives. This  means that at each potential jump one ends up with a total of $12$ matching conditions. A linear combination of four eigenvectors in each region as an eigenenergy problem naively suggests, would not provide enough free constants. However, for a generalized cubic non-linear eigenvalue problem for $k_x$ one can find 12 solutions in each region (with only some of them contributing), which is sufficient to fulfill the boundary value problem.

The generalized cubic eigenvalue problem for $k_x$ therefore reads
 \begin{equation}
 \begin{aligned}
 &\left[k_x^3\tau_z\otimes\sigma_x+3ik_yk_x^2\tau_z\otimes\sigma_2-3k_yk_x\tau_z\otimes\sigma_1\right.\left.-ik_y^3\tau_z\otimes\sigma_y+(u_j-\epsilon)\mathbb{1}_2\otimes\mathbb{1}_2\right]\psi=0.
 \end{aligned}
 \end{equation}
 
It is customary for 2D tunneling problems to choose $k_y=k\sin\theta$ to use the notion of incident angle and express the energy in terms of wavevector amplitude $k$ as $\epsilon_j=\pm k^3-u_j$. Doing so, one finds the following six $k_x$ eigenvalues that are each two-fold degenerate 
\begin{equation} 
\begin{aligned}
k_{x,1}^+&=& - k_{x,1}^- &=k \cos(\theta)\\
k_{x,2}^+ &=& - k_{x,2}^-&= - \frac{k}{\sqrt{2}}|f(\theta)|^{1/2} e^{i \gamma(\theta)}\\
k_{x,3}^+ &=& - k_{x,3}^-&= - \frac{k}{\sqrt{2}}|f(\theta)|^{1/2} e^{-i \gamma(\theta)},
\label{eq:k1}
\end{aligned}
\end{equation}
where $\gamma\left(\theta\right)=\frac{1}{2} \arg\left[f\left(\theta\right)\right]$ and $f\left(\theta\right)=\cos\left(2 \theta\right) + i \sqrt{3} - 2$, with  $\sin\left[\gamma\left(\theta\right)\right]>0$.
The wavefunctions therefore can be categorized as some of them being proportional to propagating plane waves $e^{\pm i k_{x,1}^{+} x}$ while others $e^{\pm ik_{x,(2,3)}^{+} x}$ have exponentially growing or decaying amplitudes.

To build the solution of our problem we need to determine which of the $k_{x,i}^\pm$ are compatible with the boundary conditions. For simplicity, we only consider positive energies, $\epsilon>0$.

In \textbf{region 1} ($x<-\frac{L}{2}$) we need to have finite $\psi$ as $x\to-\infty$. We realize that $e^{ik_{x,1}^\pm x}$ is a propagating wave and therefore is allowed. Next we see that $\left|e^{ik_{x,2}^{-}x}\right|\to \infty$ and $\left|e^{ik_{x,3}^{+}x}\right|\to\infty$ as $x$ tends to $- \infty$ so this state is not allowed physically. Lastly, we see that $e^{ik_{x,3}^-x}$ and $e^{ik_{x,2}^+x}$  correspond to  solutions that shrink exponentially as $x$ tends to $- \infty$ and hence are physically allowed. Therefore we have 4 physically allowed $k_x$ eigenvalues in this region, which corresponds to 8 wavefunctions. A linear combination of all these 8 contributions can be used as an appropriate ansatz
such that
\begin{equation}
\begin{aligned}
\Phi_{1}(x)&=
a_1 e_{k,1}^+\begin{pmatrix}
\chi_{k,1}^+ \\ 1 \\ 0 \\ 0
\end{pmatrix}+ a_2 e_{1,k}^+\begin{pmatrix}
0 \\ 0 \\ - \chi_{k,1}^+ \\ 1
\end{pmatrix} +a_3 e_{k,2}^+\begin{pmatrix}
 -\chi_{k,2}^+ \\ 1 \\ 0 \\ 0
\end{pmatrix}+ a_4 e_{k,2}^+ \begin{pmatrix}
0 \\ 0 \\ \chi_{k,2}^+ \\ 1
\end{pmatrix}+ a_5 e_{k,1}^-\begin{pmatrix}
 \chi_{k,1}^- \\ 1 \\ 0 \\ 0
\end{pmatrix}   \\
&+ a_6 e^-_{k,1}\begin{pmatrix}
0 \\ 0 \\ - \chi_{k,1}^- \\ 1
\end{pmatrix} + a_7  e^-_{k,3}\begin{pmatrix}
 -\chi_{k,3}^-\\ 1 \\ 0 \\ 0
\end{pmatrix}    + a_8 e^-_{k,3}\begin{pmatrix}
0 \\ 0 \\\chi_{k,3}^- \\ 1
\end{pmatrix} ,
\end{aligned}
 \label{eq:phi1-delta0-isotrop}
\end{equation}

where we have used the shorthand notations
\begin{equation}
    \chi_{k,n}^\pm=\frac{\left({k_{x,n}^\pm}^2+k^2\sin^2\theta\right)^{\frac{3}{2}}}{(k_{x,n}^\pm-ik\sin\theta)^3};\quad e_{k,n}^\pm=e^{i k_{x,n}^\pm x}.
    \label{eq:shorthands_chi}
\end{equation}

In \textbf{region 2}  ($-\frac{L}{2}<x<\frac{L}{2}$) the presence of the potential $u$ leads to a change in wavevector $\vect k$. To have a notation that distinguishes it from the other two regions that have the same wavevector we define $q=(u-\epsilon)^{1/3}$ as the wavevector amplitude in region 2 and $q_{x,y}$ as its components. We now want to express everything in terms of the new wavevector amplitude $q$. Momentum conservation in $y$-direction requires that $q_y= k_y=q \sin\phi$ with $\phi=\sin^{-1}\left(\frac{k}{q}\sin\theta\right)$.
Similarly to what happened in other regions, one finds multiple solutions for $q_{x}$, which are given by Eq. \eqref{eq:k1} if we replace $k\to q$ and $\theta\to\phi$.
Since \textbf{region 2} is of finite size there are no restrictions on the choice of $q_{x,n}^\pm$. Therefore an ansatz for the wavefunction has to include all 12 possible terms
\begin{equation}
\hspace{-0.2cm}
\begin{aligned}
 \label{eq:phi2-delta0-isotrop}
&\Phi_{2}(x)= b_1 e_{q,1}^+
\begin{pmatrix}
{-}\chi_{q,1}^+ \\ 1 \\ 0 \\ 0
\end{pmatrix}
+b_2 e_{q,1}^+\begin{pmatrix}
0 \\ 0 \\ \chi_{q,1}^+ \\ 1
\end{pmatrix}+  b_3 e_{q,2}^+\begin{pmatrix}
\chi_{q,2}^+ \\ 1 \\ 0 \\ 0
\end{pmatrix}+  b_4 e_{q,2}^+\begin{pmatrix}
0 \\ 0 \\ {{-}}\chi_{q,2}^+ \\ 1
\end{pmatrix}+  b_5 e_{q,3}^+\begin{pmatrix}
\chi_{q,3}^+ \\ 1 \\ 0 \\ 0
\end{pmatrix}+  b_6 e_{q,3}^+\begin{pmatrix}
0 \\ 0 \\ {{-}}\chi_{q,3}^+ \\ 1
\end{pmatrix}\\
&+  b_7 e_{q,1}^-\begin{pmatrix}
{{-}}\chi_{q,1}^- \\ 1 \\ 0 \\ 0
\end{pmatrix}+  b_8 e_{q,1}^-\begin{pmatrix}
0 \\ 0 \\ \chi_{q,1}^- \\ 1
\end{pmatrix} +  b_9 e_{q,2}^-\begin{pmatrix}
\chi_{q,2}^- \\ 1 \\ 0 \\ 0
\end{pmatrix}+  b_{10} e_{q,2}^-\begin{pmatrix}
0 \\ 0 \\ {{-}}\chi_{q,2}^- \\ 1
\end{pmatrix}  +  b_{11} e_{q,3}^-\begin{pmatrix}
\chi_{q,3}^- \\ 1 \\ 0 \\ 0
\end{pmatrix} +  b_{12} e_{q,3}^-\begin{pmatrix}
0 \\ 0 \\ {{-}}\chi_{q,3}^- \\ 1
\end{pmatrix},
\end{aligned}
\end{equation}
where we used the same shorthand notation as defined in Eq. \eqref{eq:shorthands_chi} with $k\to q$ and $\theta\to \phi$.

In \textbf{region 3} ($x>\frac{L}{2}$), only propagation away from the barrier is allowed and all terms need to be finite as $x\to\infty$. We find that three of the $k_x$ from Eq. \eqref{eq:k1} satisfy these conditions $k_{x,1}^+$, $k_{x,2}^-$ and $k_3^+$. The ansatz for the wavefunction in this region therefore takes the form
\begin{equation}
\begin{aligned}
\label{eq:phi3-delta0-isotrop} 
&\Phi_{3}(x) = t_1 e_{k,1}^+ \begin{pmatrix}
\chi_{k,1}^+ \\ 1 \\ 0 \\ 0
\end{pmatrix} + t_2 e_{k,1}^+\begin{pmatrix}
0 \\ 0 \\ - \chi_{k,1}^+ \\ 1
\end{pmatrix}+ t_3 e_{k,3}^+ \begin{pmatrix}
-\chi_{k,3}^+ \\ 1 \\ 0 \\ 0
\end{pmatrix}+  t_4 e_{k,3}^+ \begin{pmatrix}
0 \\ 0 \\ \chi_{k,3}^+ \\ 1
\end{pmatrix}+  t_5  e_{k,2}^- \begin{pmatrix}
-\chi_{k,2}^- \\ 1 \\ 0 \\ 0
\end{pmatrix} +  t_6  e_{k,2}^- \begin{pmatrix}
0 \\ 0 \\ \chi_{k,2}^- \\ 1
\end{pmatrix}
\end{aligned}.
\end{equation}
We can now determine the transmission coefficients $t_i$, the coefficients $a_i$ and $b_i$ from the requirement that the wavefunctions and their first and second derivatives in different regions are continuous at the two boundaries $x = - \frac{L}{2}$ and $x = \frac{L}{2}$. Continuity of the spinor wavefunctions and their first and second derivatives at each junction interface reads as follows
 \begin{equation}
 \begin{aligned}
 & \Phi_1\left(- \frac{L}{2}\right) = \Phi_2\left(- \frac{L}{2}\right); \quad \Phi_2\left(\frac{L}{2}\right) = \Phi_3\left(\frac{L}{2}\right)\\
 &{\partial}_x\Phi_1(x)|_{x=- \frac{L}{2}}={\partial}_x\Phi_2(x)|_{x=- \frac{L}{2}}\\ 
 &{\partial}_x\Phi_2(x)|_{x=\frac{L}{2}}= {\partial}_ x\Phi_3(x))_{x=\frac{L}{2}} \\
  & \partial^2_{x}\Phi_1(x)|_{x=- \frac{L}{2}}= \partial^2_{x}\Phi_2(x)|_{x=- \frac{L}{2}}\\ &\partial^2_{x}\Phi_2(x)|_{x=\frac{L}{2}}= \partial^2_{x}\Phi_3(x)|_{x=\frac{L}{2}}.
 \end{aligned}
 \label{eq:boundary_conditions}
 \end{equation}
To avoid lengthy and cumbersome expressions, we do not present the resulting system of equations. We will also not include such expressions in later sections.
 
Next, one has to compute transmission probabilities, which can be done through currents. The probability current density in $x$-direction for a state $\psi$ can be computed as follows
  \begin{equation} 
  J_x=
  \psi^{\dagger} \frac{\partial H(k)}{\partial k_x} \psi,
  \label{eq:density}
  \end{equation}
where 
\begin{equation}
\frac{\partial H(k)}{\partial k_x}=3\tau_z\otimes \left[(k_x^2-k_y^2)\sigma_x-2k_xk_y\sigma_y\right].
\end{equation}  
We may therefore use Eq. \eqref{eq:density} to compute explicitly the incident current from the wave coming from $x=-\infty$ as
   \begin{equation}
       J_x^{\sf inc}=6 k_{x,1}^+ \left[(k_{x,1}^+)^2+(k \text{sin}\theta)^2\right]^{\frac{1}{2}},
   \end{equation}
and for the transmitted wave in the asymptotic limit $x\to + \infty$ as
   \begin{equation}
       J_x^{\sf tra_{1,2}}=6 k_{x,1}^+ \left[(k_{x,1}^+)^2+(k \text{sin}\theta)^2\right]^{\frac{1}{2}}\left|t_{1,2}\right|
   \end{equation}
The transmission coefficient in this limit is then defined as
 \begin{equation}
 T=\frac{|J_x^{\sf tra}|}{|J_x^{\sf inc}|},
\end{equation}  
which allows us to compute transmission coefficient
  \begin{equation}
  T_{1,2}=\left|t_{1,2}\right|^2.
  \end{equation}
These steps allow us to compute transmission coefficients numerically.
\hiddensubsubsection{Numerical Results}
Next we present our results for this simplified version of the Hamiltonian.
As discussed previously we choose our incoming wave to have a wavevector $\vect k=k(\cos\theta,\sin\theta)$. This means that we can either set $a_1=1$ or $a_2=1$ to have a wave with these properties. Because the two blocks in the Hamiltonian differ only in sign and do not mix we find that the transmission coefficients for these cases fulfill the condition $T_1=T_2$. Therefore we only present plots for $T_1$.

\begin{figure}[H]
\centering
\subfloat[][For a potential barrier of width $L=2$ and $u_0=2$ at normal incidence $\theta=0$.]{\includegraphics[width=0.49\linewidth]{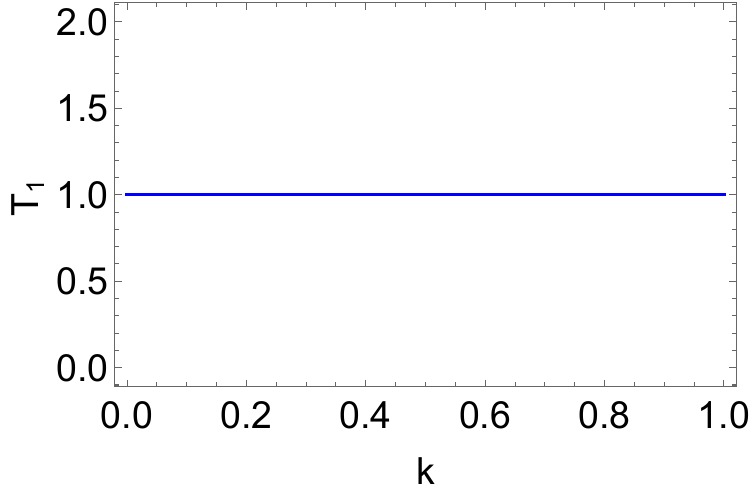}\label{fig:T1_normalinc_fixedL_rotsymm_delta0}}\hfill
\subfloat[][$u_0=4$ and small incident angle $\theta=0.1$ rad. The results were plotted for various values of the barrier width $L=2$ (blue line), $L=5$ (red line),  $L=6$ (green line).]{\includegraphics[width=0.49\linewidth]{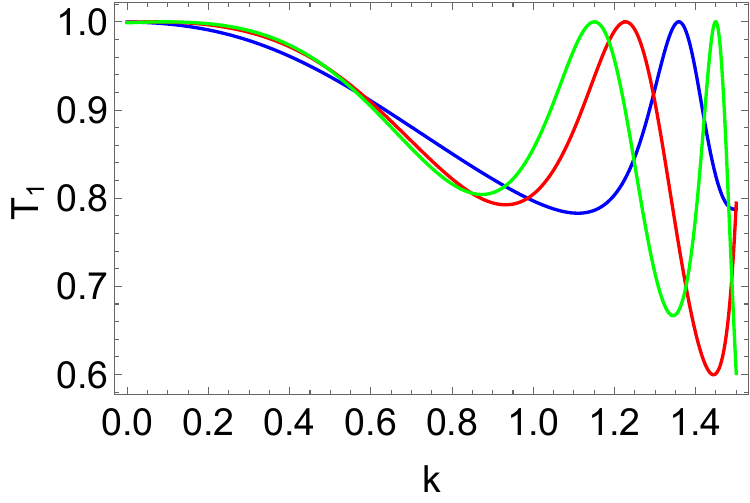}\label{fig:rotsymm_transmission_small_angle_delta0fk}}
\caption{\sf (color online) The transmission amplitude $T_1$ as a function of the incident wavevector amplitude $k$.}
\label{fig:T1_normalinc_fixedL_rotsymm_deltaAll}
\end{figure}

The transmission as a function of the wave vector $k$ is shown in Fig. \ref{fig:T1_normalinc_fixedL_rotsymm_deltaAll}.

We find that for normal incidence as seen in Fig. \ref{fig:T1_normalinc_fixedL_rotsymm_delta0} much like in the single layer graphene case, there is unit transmission up to numerical accuracy, which is commonly referred to as Klein tunneling. It is interesting to note that one block of the Hamiltonian has the same form as the Hamiltonian for ABC stacked trilayer graphene that is described in \cite{min2008electronic}. Therefore this result also applies to trilayer graphene in that approximation. One should also note that this contrasts with the result in AB bilayer graphene, where for normal incidence there is no Klein tunneling. This motivates the question of whether Klein tunneling in multilayer graphene of this stacking type is an effect that appears for an odd number of layers and disappears for an even number of layers, that is, whether it is an even-odd effect with respect to the number of layers. 

For a small incident angle keeping all other parameters fixed the result is shown in Fig. \ref{fig:rotsymm_transmission_small_angle_delta0fk} and we find that the transmission amplitude $T_1$ as a function of the  incident wavevector amplitude $k$ first decreases and then begins to oscillate. As we increase the barrier width $L$ we observe that the oscillations set in earlier.

\begin{figure}[H]
\centering
\subfloat[][Plot as a function of $L$]{\includegraphics[width=0.5\linewidth]{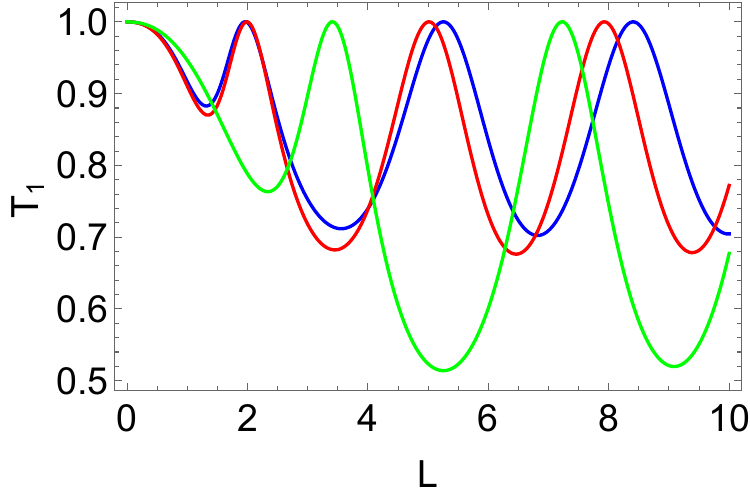}\label{fig:rotsymm_transmission_small_angle_delta0fL}}
\subfloat[][  Plot as a function of  $q_xL/\pi-b$, where $b$ is a bias.]{\includegraphics[width=0.5\linewidth]{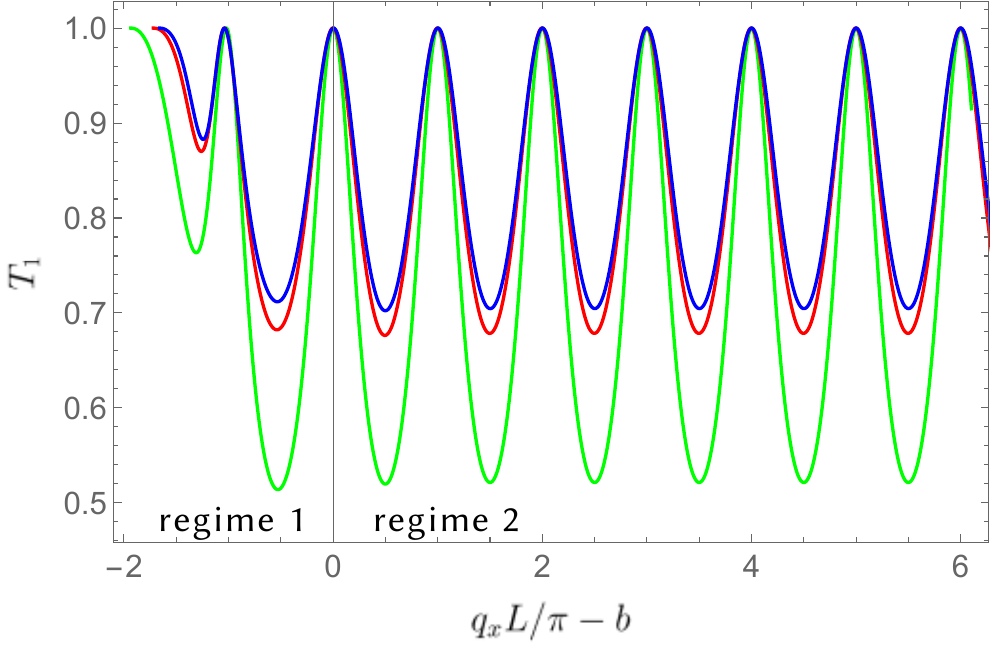}\label{fig:fig_massless_Fabry_perot}}
\caption{\sf (color online) The transmission coefficient
$T_1$ at incident angle $\theta=0.1$. We have plotted separate curves for $k=1$, $u_0=2$ (blue line), $k=1.2$, $u_0=3$ (red line) and  $k=1.5$, $u_0=4$ (green line).} \label{fig:fig_massless_Fabry_perotAll}
\end{figure}

We now turn to the transmission amplitude $T_1$ as a function of  the width of the potential barrier $L$ in Fig. \ref{fig:rotsymm_transmission_small_angle_delta0fL}.

Like in the previous plot \ref{fig:T1_normalinc_fixedL_rotsymm_delta0} we find resonances and it is not immediately obvious from Fig. \ref{fig:rotsymm_transmission_small_angle_delta0fL} why they arise.
The physical mechanism behind oscillations can, however, easily be understood if we choose the axis of our plot slightly differently as  $q_xL/\pi-b$ instead of $L$, where $q_x$ is the momentum of the propagating wave inside the barrier and $b$ is a bias that was introduced for each curve to obtain a more lucid result. 
First we may realize that there are two regimes. In regime~1 we do not observe much structure. This is because this regime corresponds to a not very thick barrier where both the exponentially decaying and oscillatory modes $q_{x,i}$ penetrate the barrier. This makes them equally important and it is difficult to disentangle their effects. It is important to note that the length of this regime also depends on the details of the barrier and incoming wave. This is why it was necessary to introduce a bias $b$ for each of the curves to render the plot for regime 2 more lucid. In regime 2 we find that the resonances fulfill $(q_x^i-q_x^{i+1})L=\pi$, where $q_x^i$ labels the $i$-th wave maximum. This requirement for constructive interference inside the barrier is commonly referred to as a Fabry-P\'erot type resonance condition.

\begin{figure}[H]
\centering
\includegraphics[width=0.5\linewidth]{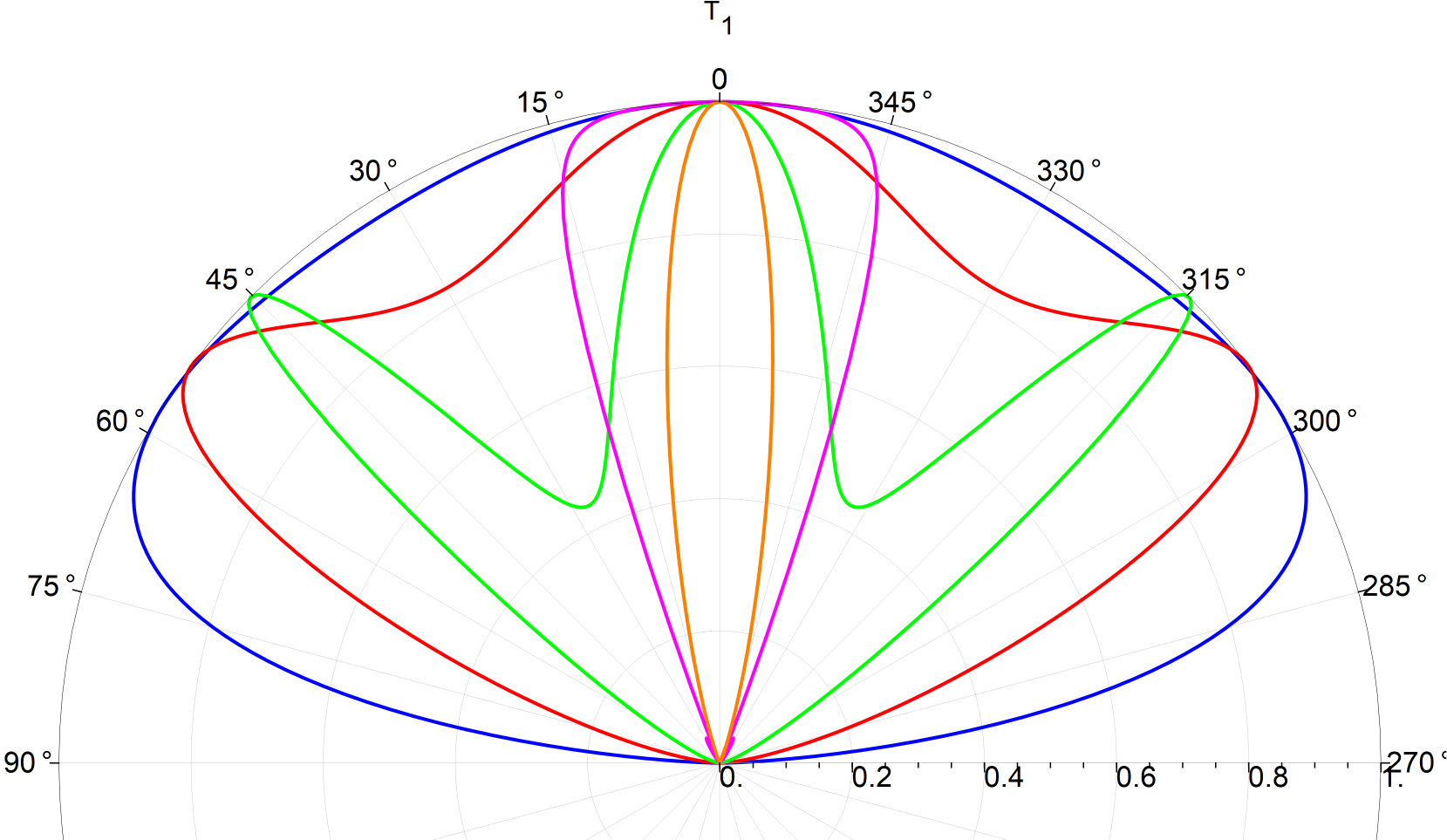}
\caption{\sf (color online)  The transmission coefficient $T_1$ as a function of the incident angle $\theta$ for $u_0=2$ and $k=1$ and for five different potential barrier widths  $L=0.2$ (blue line), 0.5 (red line), 1 (green line), 2 (magenta line), 3 (orange line).}
  \label{fig:rotsymm_transmission_small_angle_delta0_angledistr}
\end{figure}

Finally in Fig. \ref{fig:rotsymm_transmission_small_angle_delta0_angledistr} we obtain a polar plot of the transmission as a function of the incident angle $\theta$.

We find that the results are similar to those obtained for single layer graphene but are fundamentally different from those of AB bilayer graphene that shows anti-Klein tunneling at normal incidence. We have also numerically computed $\left.\frac{d^2T_1}{d\theta^2}\right|_{\theta=0}$ and thereby  found that $T_1-1\propto \theta^2+\mathcal{O}(\theta^4)$. The Klein tunneling for this case is therefore just as pronounced as in the case of graphene and not fundamentally stronger like in one of the cases we will consider later.


\subsection{Isotropic \texorpdfstring{$(k_z\neq 0)$}{(kz=0)} limit}
\label{sec:kzNotZeroIsotropic}

In this section we will discuss the case of a cubic Dirac 
semi-metal that is still isotropic in the plane but has a large enough extension in $z$-direction to allow for quantized non-zero momenta $k_z$. We may define another unitless quantity $\delta=k_z/2$, that can be interpreted as a mass-like term in the original Hamiltonian, as we will see shortly. This is similar to how in Kaluza-Klein theory with an additional periodic space direction the boundary conditions lead to a quantized momentum associated with that direction that is interpreted as a mass term. Similarly, this quantized momentum can then be interpreted as a mass term as we will see.
In this case the Hamiltonian takes the form 
 \begin{equation}
  H(k) = \left(\begin{array}{cccc} u+\delta& \hat{k}_{+}^3&0&0\\ \hat{k}_{-}^3&u-\delta&0&0\\0&0&u-\delta&- \hat{k}_{+}^3\\0&0&-\hat{k}_{-}^3&u+\delta\\\end{array}\right).
\end{equation}
The eigenvalue equation $H(k)\Psi = \epsilon \Psi$ can be solved in the different regions of space and leads to the eigenvalues 
\begin{equation}
  \epsilon_j = s_j \sqrt{k^6+\delta^2}+u_j,\qquad 
 \end{equation}
where each eigenvalue is twice degenerate, the sign function $s_j = \text{sgn}(\epsilon_{j}-u_j)$ and $j=1,2,3$ is labeling the three spatial regions.
 
\begin{figure}[H]
\centering
\includegraphics[width=0.5\linewidth]{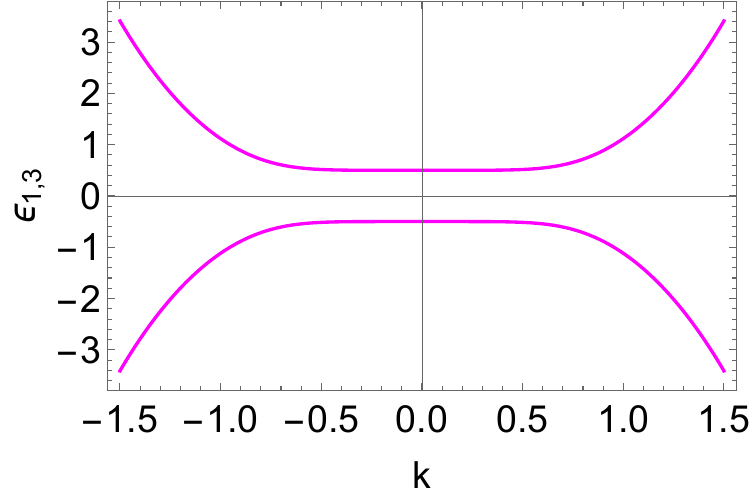}~~\includegraphics[width=0.5\linewidth]{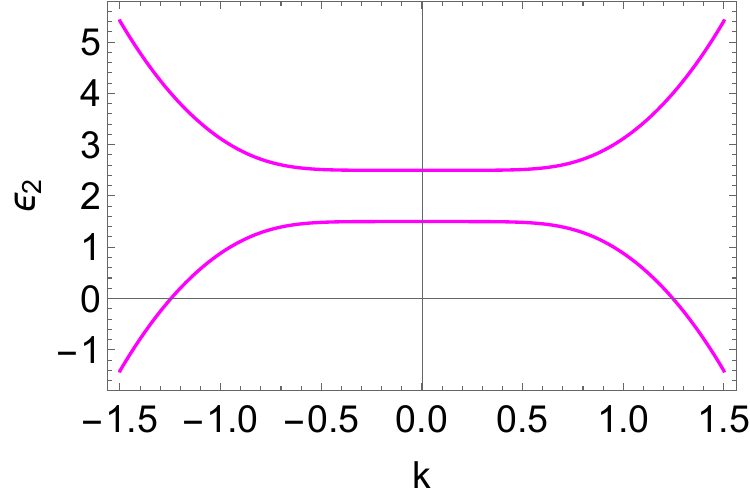}
\caption{\sf (color online) {The dispersion relation in the three different spatial regions as a function of
the incident wavevector amplitude $k$
for the value of energy gap $\delta=0.5$. The left panel shows regions 1 and 3 and  the right panel shows region 2 with $u_0=2$.  \label{fig:spectrum_gapped_isotropic}}}
\end{figure}

The band structure for the different spatial regions is shown in Fig. \ref{fig:spectrum_gapped_isotropic}.

We find that, much like a mass term, $\delta$ leads to a gap in the energy spectrum. However, it differs from the dispersion of relativistic particles or electrons in a single layer transition metal dichalcogenide \cite{Xiao_2012} in that the dispersion near $k=0$ is more flat - we quantify this to be $\epsilon\propto k^6$ near $k=0$ unlike the relativistic  case where $\epsilon\propto k^2$. 

Much like in the previous section it is advantageous to recast the problem as a cubic generalized eigenvalue problem for $k_x$
\begin{equation}
 \begin{aligned}
 &\left[k_x^3\tau_z\otimes\sigma_x+3ik_yk_x^2\tau_z\otimes\sigma_y-3k_yk_x\tau_z\otimes\sigma_1\right.\left.-ik_y^3\tau_z\otimes\sigma_y+(u_j-\epsilon)\mathbb{1}_2\otimes\mathbb{1}_2+\delta \tau_z\otimes\sigma_z\right]\psi=0,
 \end{aligned}
 \end{equation}
where $j=1,2,3$ labels the different spatial regions. With the wavevector amplitude 
$ k=(\epsilon^2-\delta^2)^\frac{1}{6}$ we find that the solutions for $k_x$ are still given by Eq. \eqref{eq:k1}.

For brevity we don't repeat the analysis of wavefunctions for the different regions because it is almost the same as in the massless case. The results are given in appendix \ref{app:wavefunctions_mass_term}.

Like previously we may now use Eq. \eqref{eq:boundary_conditions} and set $a_1=a_2=1$ to solve the remainder of the problem numerically.

\begin{figure}[H]
\centering
\includegraphics[width=0.5\linewidth]{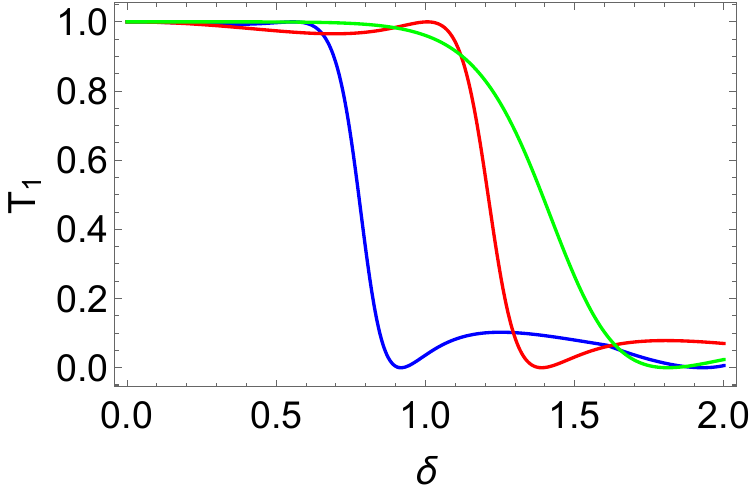}
\caption{\sf (color online) The transmission coefficient $T_1$  for normal incidence $\theta=0$  as a function energy gap  $\delta$ for a barrier width $L=2$.  We chose other values $u_0=3.5$, $k=1$ (blue line), $u_0=5$, $k=1.2$ (red line) and $u_0=7$, $k=1.5$ (green line). }
\label{fig:deltaneq0smallangle_funcdelta}
\end{figure}

\hiddensubsubsection{Numerical Results}
Next we present our numerical results for the transmission amplitudes $T_{1,2}$. Like what we did previously, we will restrict our discussion to $T_1$ because $T_{1}=T_2$ by symmetry and set $a_1=1$ to determine the properties of the incoming wave.
Since we have added a new mass term $\delta$ to the Hamiltonian, we will first and foremost be concerned with its influence. The result for a barrier width $L=2$ and normal incidence are presented in Fig. \ref{fig:deltaneq0smallangle_funcdelta}.

We find that the mass-like term $\delta$ suppresses tunneling like one would expect from the semi-classical intuition that mass suppresses quantum phenomena like tunneling. We also find that there are certain values of $\delta$ where the transmission probability despite increasing the mass-like term rises again and for some specific values even reaches unity again. This is likely due to Fabry-P\'erot type interference effects inside the barrier, which we will confirm shortly.

\begin{figure}[H]
\centering
\subfloat[][Incidence at $\theta=0$ rad.]{\includegraphics[width=0.49\linewidth]{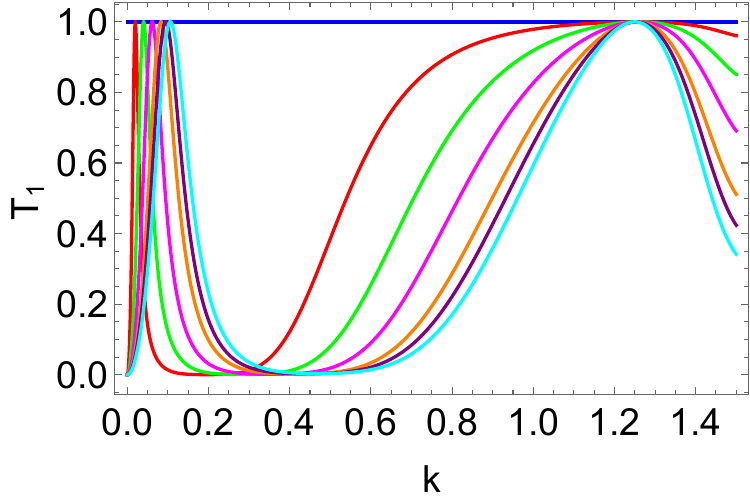}}
\subfloat[][Incidence at $\theta=0.1$ rad.]{\includegraphics[width=0.49\linewidth]{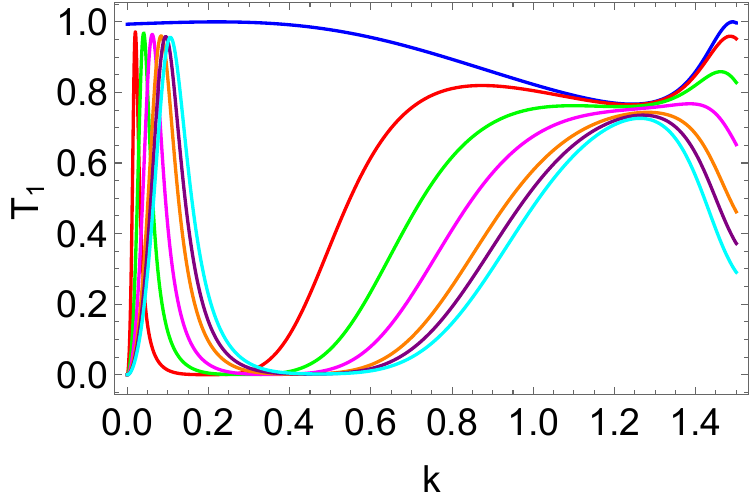}}
\caption{\sf (color online) The transmission amplitude $T_1$ as a function of the incident wavevector amplitude $k$  for $u_0=4$ and barrier width $L=2$. We have plotted separate curves for different energy gaps $\delta=0$ (blue line), $\delta=0.1$ (red line), $\delta=0.2$ (green line), $\delta=0.3$ (magenta line), $\delta=0.4$ (orange line), $\delta=0.45$ (purple line) and $\delta=0.5$ (cyan line).}
\label{fig:deltaneq0smallangle_funck}
\end{figure}

To gain further intuition let us now study $T_1$ as a function of the incident 
wavevector amplitude $k$ as shown in Fig. \ref{fig:deltaneq0smallangle_funck} .
We find that the inclusion of the mass terms leads to a gap in the transmission amplitudes in the region of $0.2 \lesssim k\lesssim 0.8$. This transmission gap is again best explained semi-classically. A particle with increasing mass will behave increasingly classical. Classical particles cannot tunnel through a barrier. Therefore with increasing mass one observes a transmission gap. We also see that for incidence that is not normal the transmission amplitude does not reach unity again in the massive case but the resonances appear at approximately the same spots as the massless case. They are therefore related to the same physical phenomenon.

\begin{figure}[H]
\centering
\subfloat[][Plot as function of $L$]{\includegraphics[width=0.5\linewidth]{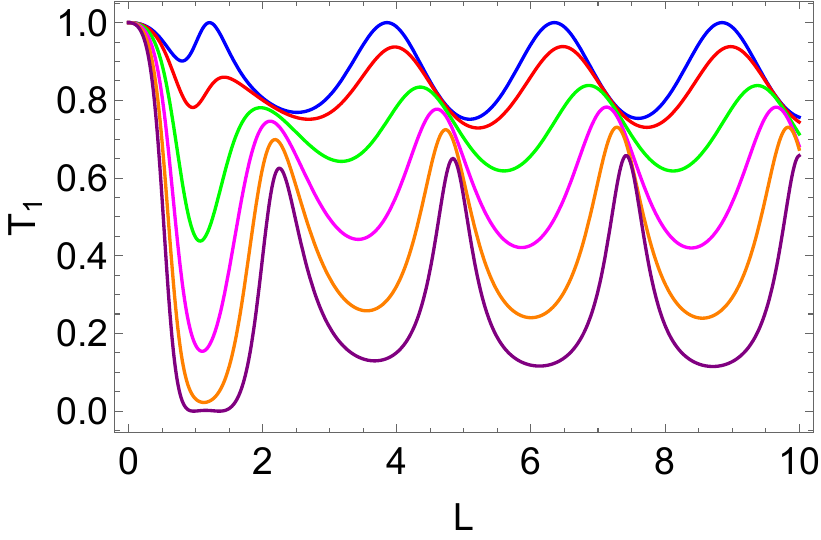}\label{fig:deltaneq0smallangle_funcL}}
\subfloat[][ Plot as a function of  $q_xL/\pi-b$]{\includegraphics[width=0.5\linewidth]{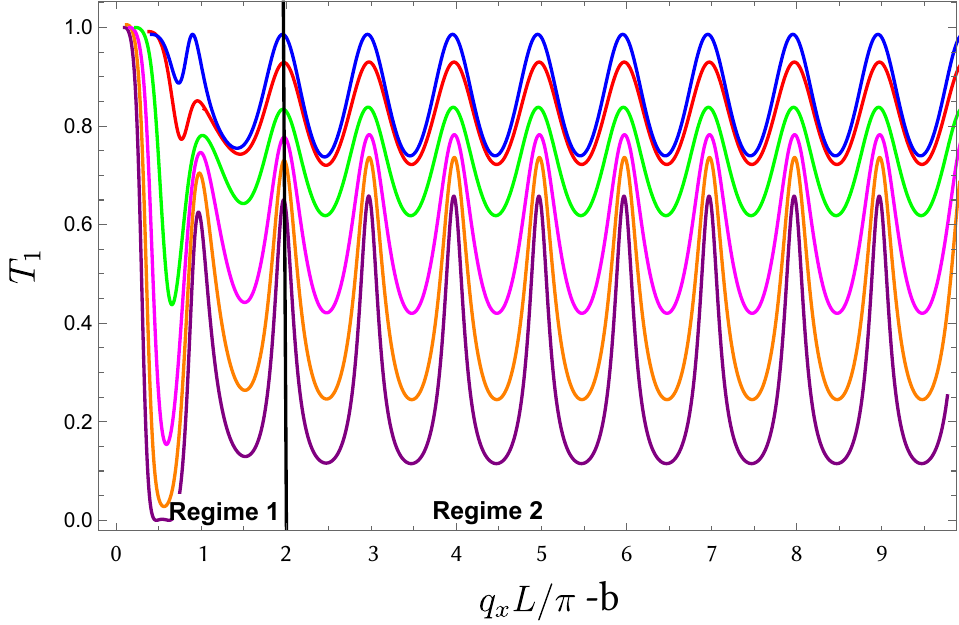}\label{fig:fig_mass_Fabry_perot}}
\caption{\sf (color online) The transmission amplitude $T_1$  as a function of barrier width for incident wavevector amplitude $k=1$, $u_0=3$ and $\theta=0.1$ rad. We  have plotted separate curves for energy gaps $\delta=0$ (blue line), $0.1$ (red line), $0.2$ (green line), $0.3$ (magenta line), $0.4$ (orange line) and $\delta=0.5$ (purple line).} \label{fig:deltaneq0smallangle_funcLFabry}
\end{figure}

To understand this phenomenon better and relate it to the Fabry-P\'erot interferrometer we again plot the transmission amplitude $T_1$ as a function of the width of the potential barrier $L$ in Fig. \ref{fig:deltaneq0smallangle_funcLFabry}. 
We first observe that oscillations as a function of barrier thickness $L$ become increasingly sharp peaks as we increase the magnitude of the mass term, $\delta$. We observe a suppression of tunneling except at certain specific resonant values. 

To understand the physical mechanism underlying these resonances  in Fig. \ref{fig:fig_mass_Fabry_perot} we plot the transmission coefficient as a function of $q_xL/\pi-b$, where $b$ is a bias that was introduced to obtain a more lucid plot.

Like in the mass-less case we find again that there are two regimes. Regime 1 does not have much structure. This is because this regime corresponds to a thin barrier where both the exponentially decaying and oscillatory modes $q_{x,i}$ contribute equally inside the barrier.  In regime 2 we find that the resonances fulfill $(q_x^i-q_x^{i+1})L=\pi$, where $q_x^i$ labels the $i$-th wave maximum. This is similar to previous section and this requirement is called the Fabry-P\'erot resonance condition.

\begin{figure}[H]
 \centering
\includegraphics[width=0.6\linewidth]{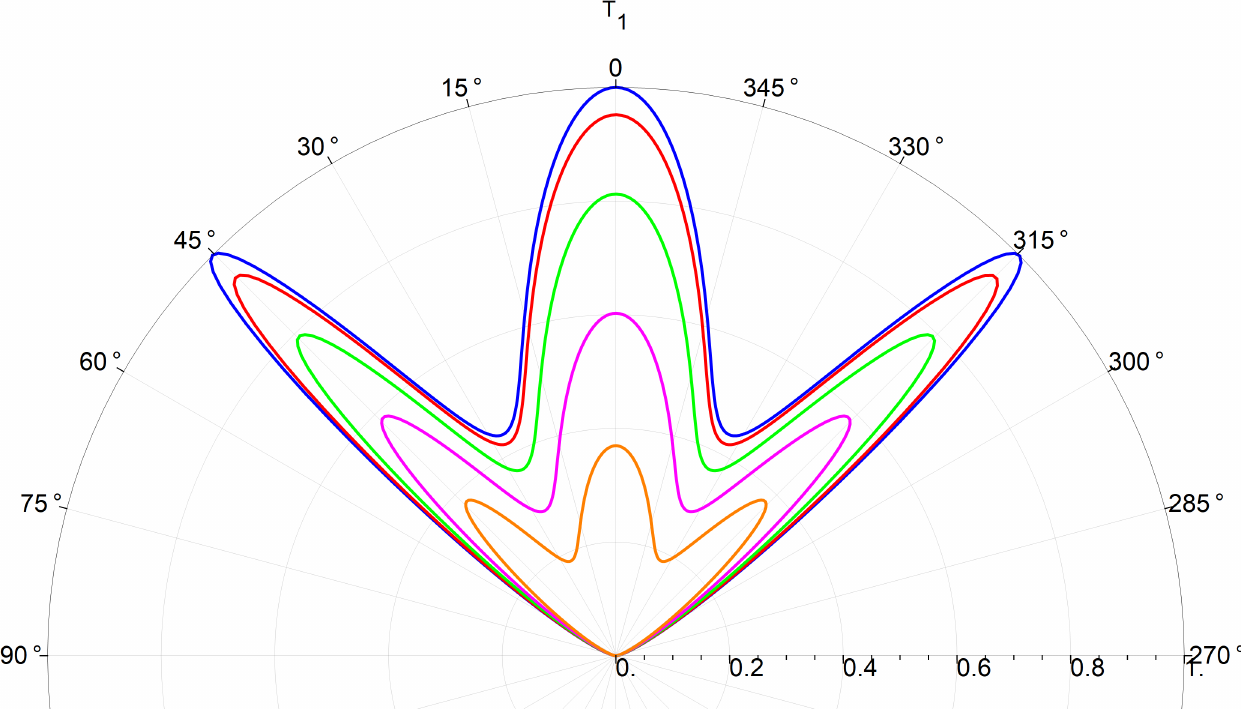}
\caption{\sf (color online)  The angular dependence of the transmission coefficient $T_1$ for  $u_0=2$, $k=1$ and barrier width $L=1$ for five values of the energy gap $\delta=0$ (blue line), $0.1$ (red line), $0.2$ (green line), $0.3$ (magenta line),  $0.4$ (orange line). } 
\label{fig:deltaneq0smallangle_functheta}
\end{figure}

In Fig. \ref{fig:deltaneq0smallangle_functheta} we show polar plots of the transmission coefficient $T_1$ as a function of  the incident angle $\theta$.

For the chosen parameters we find that the main effect of the mass-like term, $\delta$, is to weaken the strength of the barrier transmission. This is in full agreement with the semi-classical intuition one has that a heavier particle is "less quantum" and therefore should be less susceptible to exhibit quantum effects such as tunneling.

\subsection{Anisotropic 2D \texorpdfstring{$(k_z=0)$}{(kz=0)} limit}
\label{Sec:anisotropkz0}

\begin{figure}[H]
\centering
\subfloat[][Contour plot of the positive energy band.]{\includegraphics[width=0.5\linewidth]{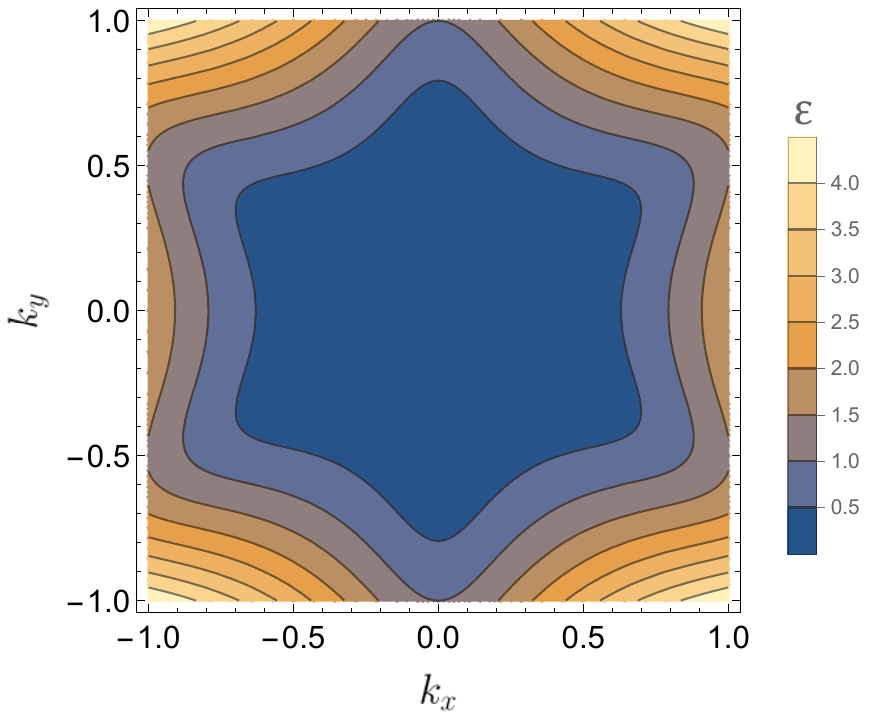}}
\subfloat[][Slice at $k_y=0$ for both energy solutions]{\includegraphics[width=0.5\linewidth]{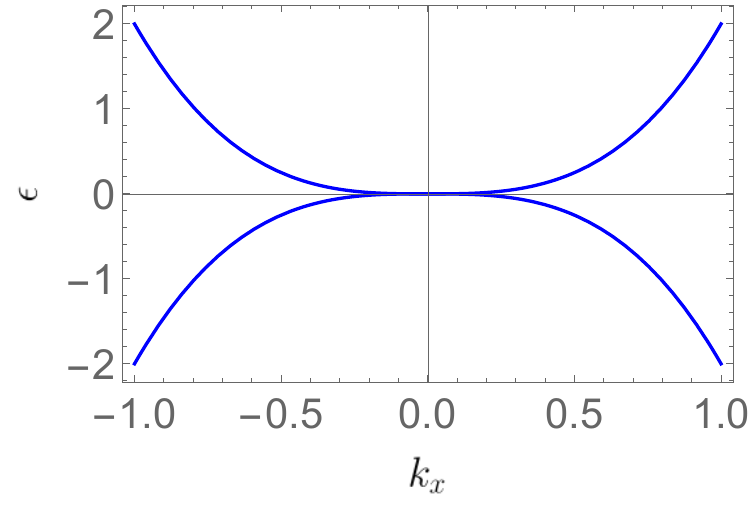}}
\caption{\sf (color online) Plots of the dispersion relation in region 1 for $\lambda=2$. } \label{fig:Energyplot_anisotropdelta0}
\end{figure}

In this section we will consider the case that is anisotropic in the $x$-$y$ plane, where $v_y= \lambda v_x=\lambda v$. We consider again the case of thin material with $k_z=0$. Much like in the previous cases it is advantageous to define unitless quantities via Eq. \eqref{eq:unitless_en}.
The Hamiltonian of the system  is given by
 \begin{equation}
  H(k) = \left(\begin{array}{cccc} 0& \hat{k}_{1}&0&0\\ \hat{k}_{1}^\dagger&0&0&0\\0&0&0&- \hat{k}_{1}\\0&0&-\hat{k}_{1}^\dagger&0\\\end{array}\right)+u_j\mathbb{1}_4,
\end{equation}
where
\begin{equation}
\hat{k}_{1}=\frac{\lambda+1}{2} \hat{k}_{+}^3+\frac{\lambda-1}{2} \hat{k}_{-}^3.
\end{equation}
For this anisotropic Hamiltonian we find that the energy eigenvalues 
\begin{equation}
  \epsilon_j = \pm \sqrt{\lambda^2 k_x^2 \left(k_x^2-3 k_y^2\right)^2+\left(k_y^3-3 k_x^2 k_y\right)^2}+ u_j,
 \end{equation}
also have an anisotropic energy dispersion relation that has a sixfold rotational symmetry as shown in Fig. \ref{fig:Energyplot_anisotropdelta0}.

We find that even though rotational symmetry is broken, the system is still gap-less.

\hiddensubsubsection{Solution of the tunneling problem}
Before we investigate the tunneling problem in detail let us make an interesting observation  about $k_1=\lambda  k_x^3+3 i k_x^2 k_y-3 \lambda  k_x k_y^2-i k_y^3$. Namely we find that if we make the following transformation   
\begin{equation}
    \lambda\to\frac{1}{\lambda};\quad k_{x,y}\to \lambda^{\frac{1}{3}}k_{x,y},
    \label{eq:replacements_aniso}
\end{equation}
then up to a relabeling of Pauli matrices the roles of $k_x$ and $k_y$ in the full Hamiltonian get interchanged, that is, $k_1\to   k_x^3+3 i \lambda k_x^2 k_y-3   k_x k_y^2-i \lambda k_y^3$. This means that the replacements \eqref{eq:replacements_aniso} amounts to a hidden symmetry in parameters that allows us to use results for tunneling through a barrier in $x$-direction and without additional effort deduce results for tunneling through a barrier in $y$-direction. One just has to make proper use of the replacements \eqref{eq:replacements_aniso}. We will therefore, in our analytical treatment restrict ourselves to a potential barrier in $x$-direction but will still be able to compute results for tunneling through a barrier in $y$-direction.

\begin{figure}[H]
\centering
\includegraphics[width=0.5\linewidth]{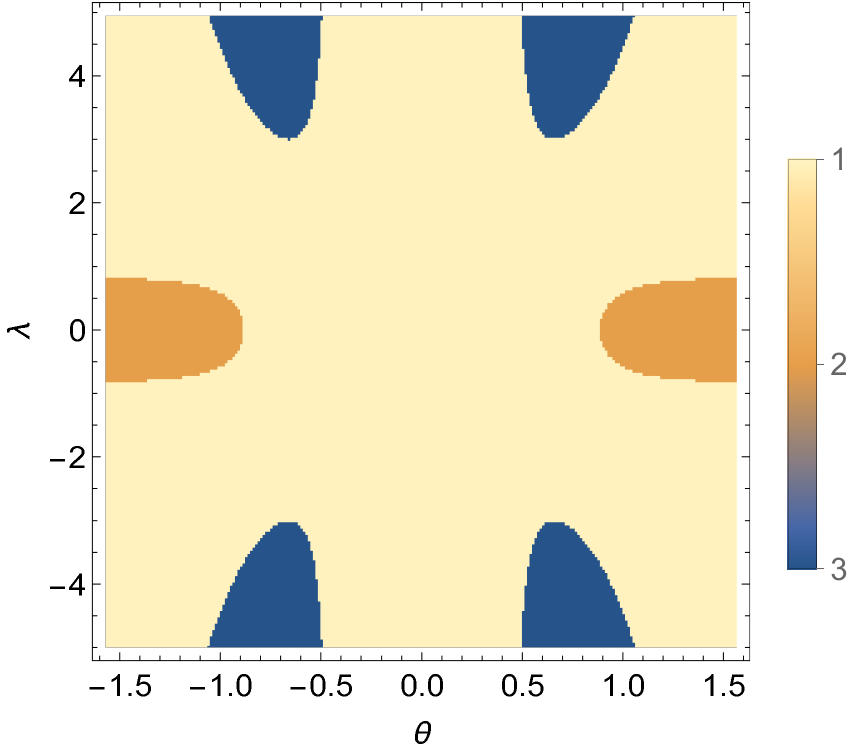}
\caption{\sf (color online) The number of real valued solutions $k_{x,i}^+$ as function of anistropy pamarameter $\lambda$ and incident angle $\theta$ (in units of rad). This can be interpreted as number of propagating modes that can be detected by a distant detector. It is a tunneling phase diagram.} \label{fig:anisotrop_lambda_theta_phase_diag}
\end{figure}

For an incoming plane wave with momentum $\vect k=(k\cos\theta,k\sin\theta)$
we find that the possible eigenenergies are given by
 \begin{equation}
 \epsilon=\pm \frac{1}{\sqrt{2}}\sqrt{k^6 \left[\left(\lambda ^2-1\right) \cos (6 \theta )+\lambda ^2+1\right]},
 \label{eq:anisotropenergy}
 \end{equation}
and each eigenvalue is twice degenerate.
We may now determine the allowed momenta $k_x$ at this energy in the different regions if we solve the non-linear generalized eigenvalue problem
\begin{equation}
\begin{aligned}
   &[ \lambda  k_x^3\tau_z\otimes\sigma_x-3  k_x^2 k_y\tau_z\otimes\sigma_y-3 \lambda  k_x k_y^2\tau_z\otimes \sigma_x+ k_y^3\tau_z\otimes \sigma_y+(u_j-\epsilon)\mathbb{1}_4]\psi=0,
   \end{aligned}
   \label{eq:anisotropkz0kxeigeneq}
\end{equation}
for $k_x$ at fixed energy $\epsilon_j$ given in equation \eqref{eq:anisotropenergy} and $k_y=k\sin\theta$.
We find that there are six possible wavevectors $k_x$ in {\bf region 1}, which are given by
\begin{equation} 
\begin{aligned}
k_{x,1}^+&= - k_{x,1}^- =k \cos(\theta)\\
k_{x,2}^+ &= - k_{x,2}^-= - \frac{k}{2 \lambda}\mathrm{sign}(\gamma_-(\theta))|f_+(\theta)|^{1/2} e^{i \gamma_-(\theta)}\\
k_{x,3}^+ &= - k_{x,3}^-= \mathrm{sign}(\gamma_+(\theta)) \frac{k}{2 \lambda}|f_-(\theta)|^{1/2} e^{i \gamma_+(\theta)},
\end{aligned}
\label{eq:anisotrop_k1}
\end{equation}
where the sign convention is chosen to allow an easier classification of complex wavevectors.

\begin{equation}
\eqfitpage{f_{\pm}\left(\theta\right)=\lambda ^2 (5-7 \cos (2 \theta ))\pm\sqrt{\frac{3}{2}} \Big[81-4 \left(\lambda ^4-28 \lambda ^2+27\right) \cos (2 \theta )+\left(27-5 \lambda ^4\right.\left.-22 \lambda ^2\right) \cos (4 \theta )+\lambda ^4-90 \lambda ^2\Big]^{\frac{1}{2}}-18 \sin ^2(\theta )},
\label{eq:anistrop_k1_f}
\end{equation}
and $\gamma_{\pm}\left(\theta\right)=\frac{1}{2} \arg\left[f_{\pm}\left(\theta\right)\right]$, with  $\sin\left[\gamma_{\pm}\left(\theta\right)\right]>0$.
The solutions for $k_{x,i}^\pm$ can be real or complex valued depending on the choice of parameters. In Fig. \ref{fig:anisotrop_lambda_theta_phase_diag}
we present a plot of many of the solutions $k_{x,i}^+$ that are real valued.

We find that there are regions with different numbers of real valued solutions. It is useful to recognize that the number of real solutions corresponds to the number of modes that are purely propagating - not exponentially decaying. They therefore can be detected by a distant detector. In this sense the figure shows a kind of \textit{tunneling phase diagram of a distant detector}. 

Because the wave vectors in each region in Fig. \ref{fig:anisotrop_lambda_theta_phase_diag} have different properties one has to be careful. Particularly, one will require a different ansatz for the wavefunction that takes this into account. In this paper we will restrict ourselves to the light brown regions with only $k_{x,1}^+$ as real valued and $k_{x,2/3}^+$ complex valued. In all further plots later in this section we will restrict our parameters so that each plot point falls into one of these regions.

The ansatz for the wavefunction in the different regions is found similarly to Sec. \ref{sec:Sol:wf:massless_isotrol} by a discussion of which wavevectors are physically allowed. For brevity we omitted this discussion from the main text and have relegated it to appendix \ref{app:Wavefunct_non_rot_symm_zero_mass}. With this ansatz for the wavefunction and by matching conditions \eqref{eq:boundary_conditions}  we may then solve the problem numerically for the different coefficients. We will now turn to these numerical results.

\hiddensubsubsection{Numerical Results}

\begin{figure}[H]
\centering
\subfloat[][Barrier in $x$-direction]{\includegraphics[width=0.5\linewidth]{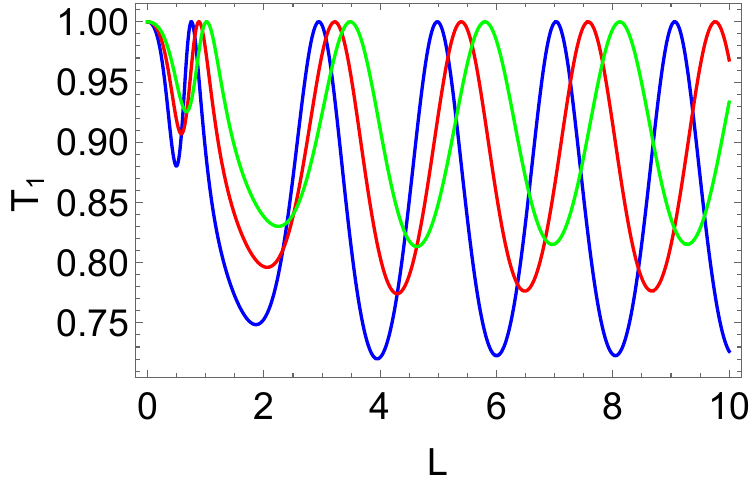}}
\subfloat[][Barrier in $y$-direction.]{\includegraphics[width=0.5\linewidth]{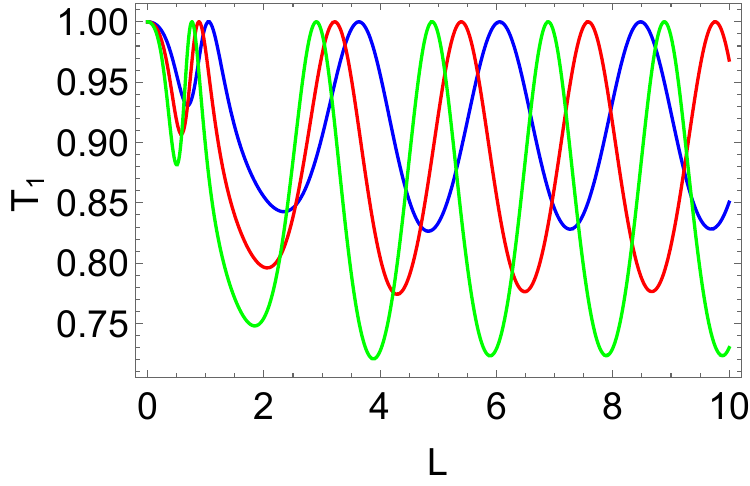}}
\caption{\sf (color online) The transmission coefficient $T_1$ as a function of the barrier thickness $L$ for the incident wavevector amplitude $k=1$, barrier height $u_0=4$, incident angle $\theta=0.1$ rad and various values of $\lambda=0.85$ (blue line), $1$ (red line), $1.15$ (green line).} \label{fig:anisortropdelta0figL}
\end{figure}

First, in Fig. \ref{fig:anisortropdelta0figL}  we present the transmission coefficient $T_1$ as a function of the width of the potential barrier $L$. We do this for two different orientations of the barrier to capture the effect of the anisotropy.

As one would expect we find that the rotational symmetry is broken for $\lambda\neq 1$. We also find that the value of transmission minima changes for different  values of $\lambda$ . This change in height as we change $\lambda$ can be attributed to the fact that the energy of the incident particle not only depends on $k$ but also on $\lambda$. This means that the curves compare tunneling of particles at different energies. The apparent symmetry between blue and green curves for the two plots is of no deeper physical relevance - it is related to the specific choice of parameters. We also find that the location of the transmission maximum changes for different $\lambda$. This is because $\lambda$ enters into $q_{x,i}$ and therefore also impacts where Fabry-P\'erot type resonances occur.

\begin{figure}[H]
\centering
\includegraphics[width=0.5\linewidth]{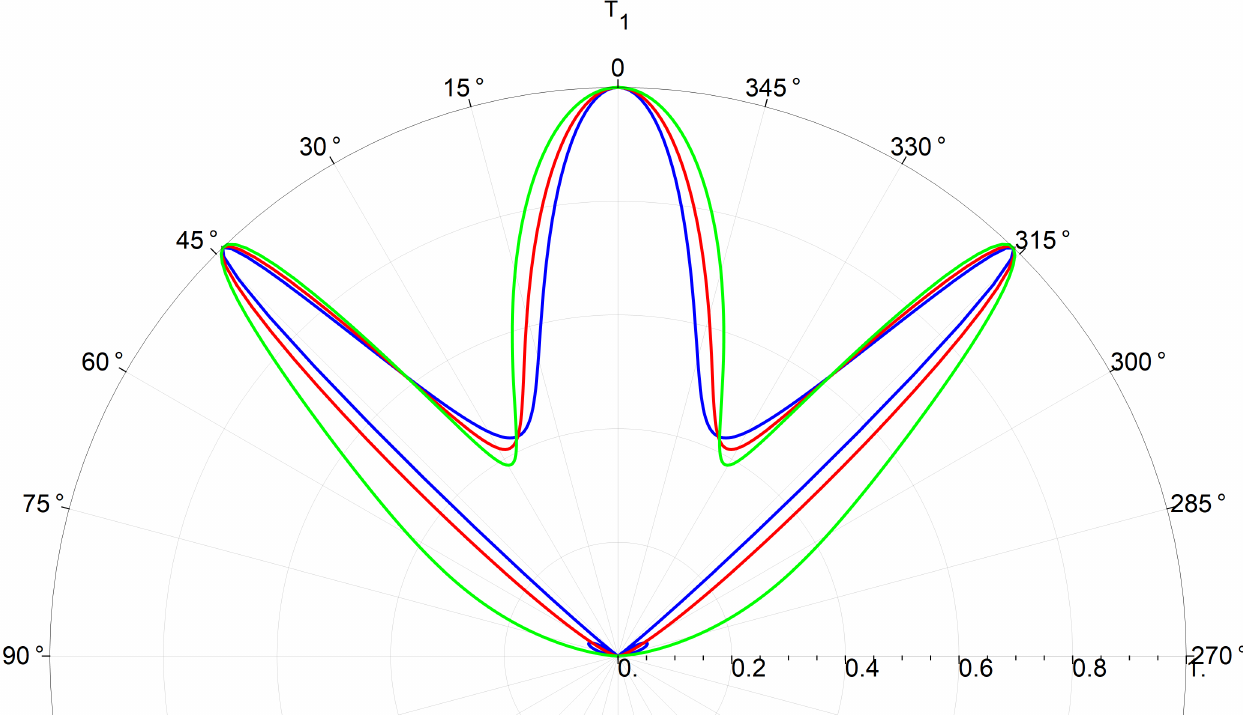},\includegraphics[width=0.5\linewidth]{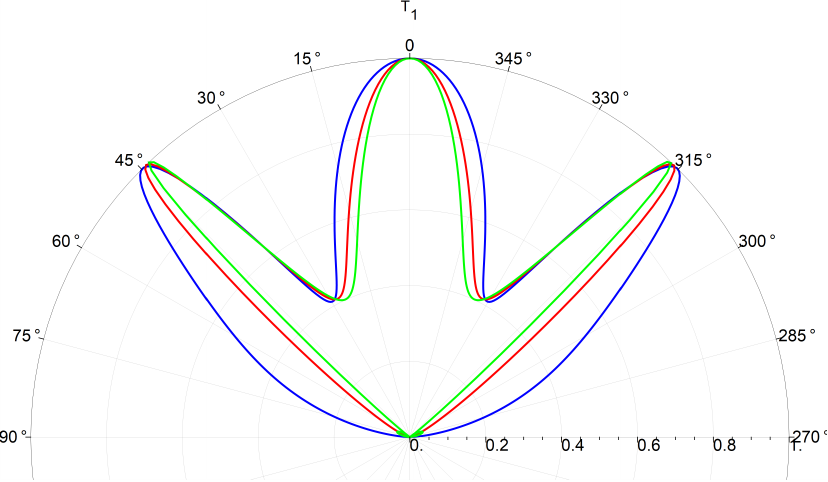}
\caption{\sf (color online) The transmission coefficient $T_1$ as a function of the incident angle $\theta$ for $k=1$, $u_0=10$, $L=0.2$ and various values of $\lambda=0.85$ (blue line), $1$ (red line), $1.15$ (green line). The left panel shows results for a barrier located in $x$-direction and the right panel for a barrier in $y$-direction.} \label{fig:anisortropdelta0figtheta}
\end{figure}

Last, we present the transmission coefficient $T_1$ as a function of the incident angle $\theta$ in Fig. \ref{fig:anisortropdelta0figtheta}.

Much like in previous plots we find that the most prominent feature is the broken rotational symmetry for $\lambda\neq 1$ - that is the results for a barrier in $x$-direction and a barrier in $y$-direction differ.

\subsection{Anisotropic \texorpdfstring{$(k_z\neq 0)$}{(kz=0)} limit}
\label{sec:anisotrop_kz_not_zero}

In this section we consider the situation where we have both anisotropy $v_y=\lambda v_x=\lambda v$ and $k_z\neq 0$. The reason for doing this is to see if there is any effect that is uniquely associated with the interplay between anisotropy parameter $\lambda$ and finite momentum $k_z$.
It is convenient to recast the problem using dimensionless energy $\epsilon=\frac{E}{2 \hbar v }$, gap $\delta=\frac{v_zk_z}{2 v}$ and potential $u_j=\frac{V_j}{2 \hbar v }$. It may then be described by the Hamiltonian
 \begin{equation}
  H(k) = \left(\begin{array}{cccc} \delta& \hat{k}_{1}&0&0\\ \hat{k}_{1}^\dagger&-\delta&0&0\\0&0&-\delta&- \hat{k}_{1}\\0&0&-\hat{k}_{1}^\dagger&\delta\\\end{array}\right),
\end{equation}
where
\begin{equation}
\hat{k}_{1}=\lambda  \hat{k}_{x}^3+3 i \hat{k}_{x}^2 \hat{k}_{y}-3 \lambda  \hat{k}_{x} \hat{k}_{y}^2-i \hat{k}_{y}^3.
\end{equation}
We find that the energies in {\bf region 1} are given by
\begin{equation}
 \label{D}
  \epsilon_j = \pm \sqrt{\lambda^2 k_x^2 \left(k_x^2-3 k_y^2\right)^2+\left(k_y^3-3 k_x^2 k_y\right)^2+\delta^2}
 \end{equation}
and are twice degenerate.

\begin{figure}[H]
  \centering
\subfloat[][Contour plot of the positive
energy band.]{\includegraphics[width=0.5\linewidth]{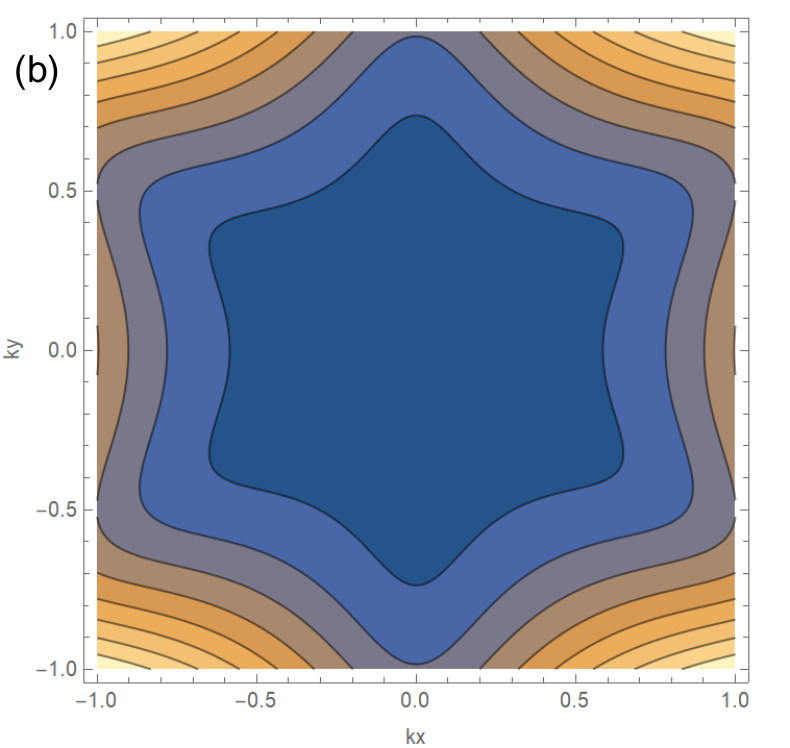}}
\subfloat[][Slice at $k_y=0$ of both energy solutions]{\includegraphics[width=0.5\linewidth]{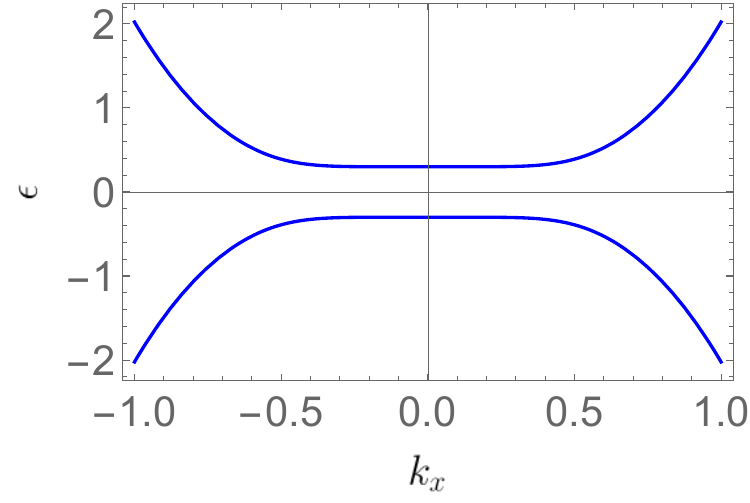}}
\caption{\sf (color online)  Plot of the dispersion in region 1 for $\lambda=2$ and  $\delta=0.3$. }
\label{f1}
\end{figure}

A plot can be seen in Fig. \ref{f1}.
We observe a sixfold symmetry and an energy gap.

We will now turn to the solution of the tunneling problem.
Here, we may consider in {\bf region 1} the case of an incoming wave with wavevector $\vect k=k(\cos\theta,\sin\theta)$, which is associated with positive energy eigenvalues given as
 \begin{equation}
 \epsilon=\frac{1}{\sqrt{2}} \sqrt{2 \delta^2 + k^6 \left[\left(\lambda ^2-1\right) \cos (6 \theta )+\lambda ^2+1\right]}.
 \end{equation}
At this energy one may determine the permitted wavevectors that can appear through interaction with the potential barrier. For that, one can interpret the eigenenergy equation as a non-linear generalized eigenvalue problem for $k_x$
\begin{equation}
\begin{aligned}
   &[ \lambda  k_x^3\tau_z\otimes\sigma_x-3  k_x^2 k_y\tau_z\otimes\sigma_y-3 \lambda  k_x k_y^2\tau_z\otimes \sigma_x+ k_y^3\tau_z\otimes \sigma_y+\delta\tau_z\otimes\sigma_z+(u_j-\epsilon)\mathbb{1}_4]\psi=0.
   \end{aligned}
   \label{eq:anisotropkznot0kxeigeneq}
\end{equation}

The possible solutions $k_x$ for regions 1 and 3 are again given by Eqs. (\ref{eq:anisotrop_k1}-\ref{eq:anistrop_k1_f}) and are therefore independent of $\delta$ but can take both complex and real values. Real valued solutions propagate to infinity, complex valued solutions due to the boundary conditions should only be included if they decay at infinity. Therefore a distant detector will only detect waves corresponding to real-valued solutions of $k_x$. Analogously, we may use Fig. \ref{fig:anisotrop_lambda_theta_phase_diag} to classify the parameter regions that have a certain number of real solutions $k_{x,i}^+$. In the following we will again restrict our discussion to light brown regions with only one of the $k_{x,i}^+$ being real valued and will make this assumption for any ansatz wavefunction.

Determining the ansatz wavefunctions in the different spatial regions follows the same approach as in Sec. \ref{sec:Sol:wf:massless_isotrol} and therefore is again relegated to an appendix \ref{app:Wavefunct_non_rot_symm_massive}.

We can now solve the problem numerically if we insert the ansatz from appendix \ref{app:Wavefunct_non_rot_symm_massive} in \eqref{eq:boundary_conditions} and solve for the coefficients. 

\begin{figure}[H]
\centering
\subfloat[][Barrier in $x$-direction]{\includegraphics[width=0.5\linewidth]{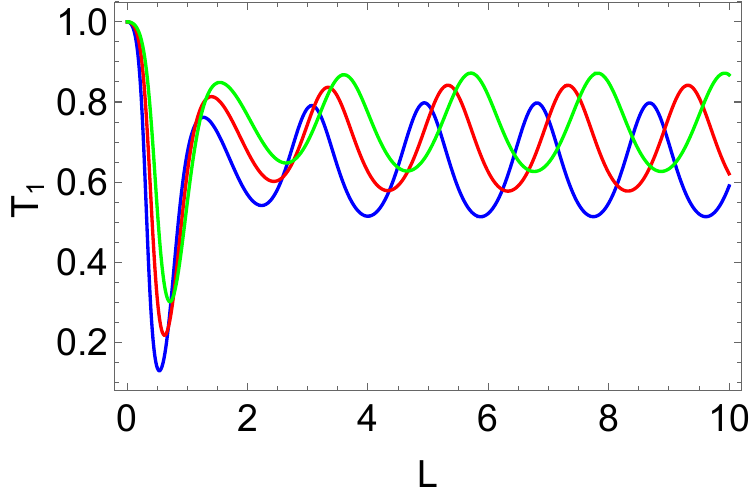}}\subfloat[][Barrier in $y$-direction]{\includegraphics[width=0.5\linewidth]{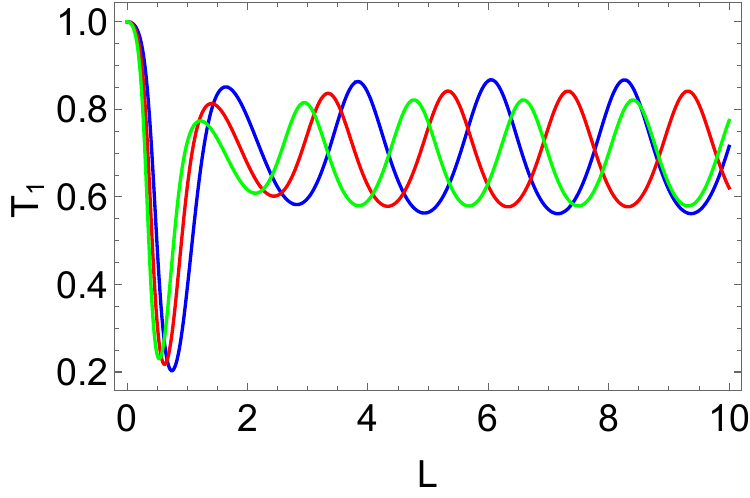}}
\caption{\sf (color online) The transmission coefficient
$T_1$ as a function of the barrier thickness $L$ for the incident wavevector amplitude
$k=1$, barrier height $u_0 =5$, incident angle $\theta=0.1$ rad and mass term $\delta=0.3$. We plotted various
values of $\lambda=0.85$ (blue line), $1$ (red line), $1.15$ (green line). } \label{fig:anisortropdeltaNot0figL}
\end{figure}

\hiddensubsubsection{Numerical Results}
We will again set $a_1=1$ to choose the incoming wave as one with wavevector $\vect k=(\cos\theta,\sin\theta)k$. This allows us to determine the tunneling amplitude which is then used to compute the transmission amplitude $T_1$ shown in Fig. \ref{fig:anisortropdeltaNot0figL} as a function of the width of the potential barrier $L$.

We find that transmission probabilities get suppressed by the addition of the mass-like term $\delta$ as one would expect from previous sections and semi-classical intuition which dictates that a mass terms suppresses quantum effects such as tunneling. Addition of the anisotropy $\lambda\neq 1$ breaks rotational symmetry, however, there does not appear to give rise to any special effect that is unique to the coexistence of both anisotropy and mass-like term. The oscillations are again due to a Fabry-Perot-type resonance condition similar to previous sections.

\begin{figure}[H]
\centering
\subfloat[][Barrier in $x$-direction]{\includegraphics[width=0.5\linewidth]{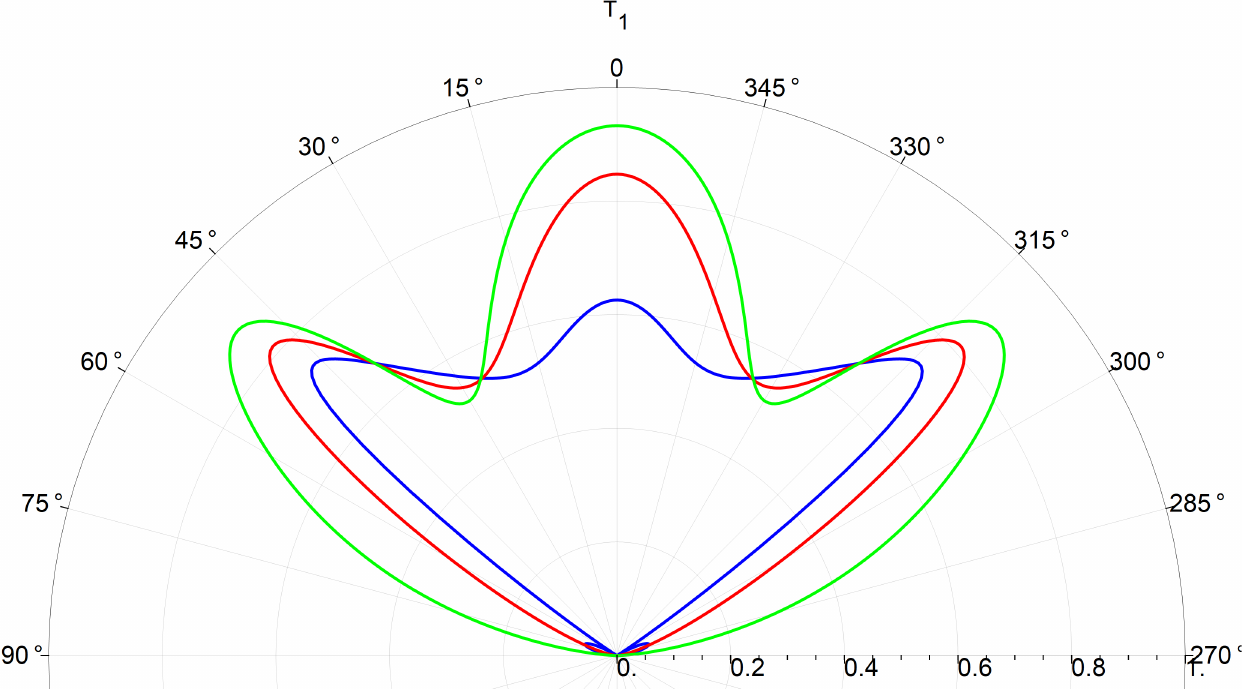}}
\subfloat[][Barrier in $y$-direction]{\includegraphics[width=0.5\linewidth]{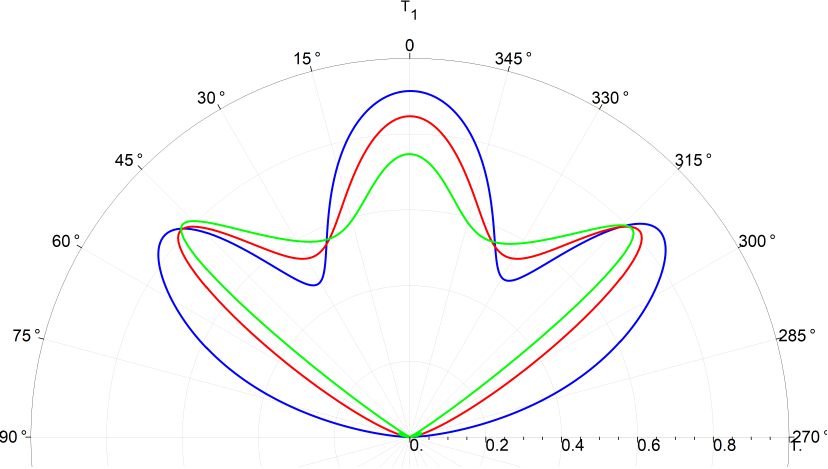}}
\caption{\sf (color online) The transmission coefficient
$T_1$ as a function of the incident angle $\theta$ for a barrier width $L=0.5$, barrier height $u_0 =3$, wavevector amplitude $k=1$, $\delta=0.3$ and various values of $\lambda=0.85$ (blue line), $1$ (red line), $1.15$ (green line). } \label{fig:anisortropdeltaNot0figtheta}
\end{figure}

Finally, in Fig. \ref{fig:anisortropdeltaNot0figtheta} we show the transmission coefficient $T_1$ as a function the incident angle $\theta$.

We observe that much like in all other plots the mass-like term $\delta$ leads to smaller maximum values of transmission coefficient $T_1$ while the addition of the anisotropy $\lambda\neq 1$ breaks the rotational symmetry. 
We were unable to find an effect that uniquely depends on the interplay between $\lambda$ and $\delta$. It seems that if there exists such an effect then it might likely occur at values of $\lambda$ and $\theta$ that give rise to more than one real solution. That is they may occur in regions of Fig. \ref{fig:anisotrop_lambda_theta_phase_diag} that are dark brown or blue.


\subsection{Bulk transmission through a potential along z-direction}
\label{sec:tunneling_z_dir}

\begin{figure}[H]
\begin{tabular}{cc}
    \begin{minipage}{0.5\textwidth} \subfloat[][Contour plot of the
positive energy band]{\includegraphics[width=1\textwidth]{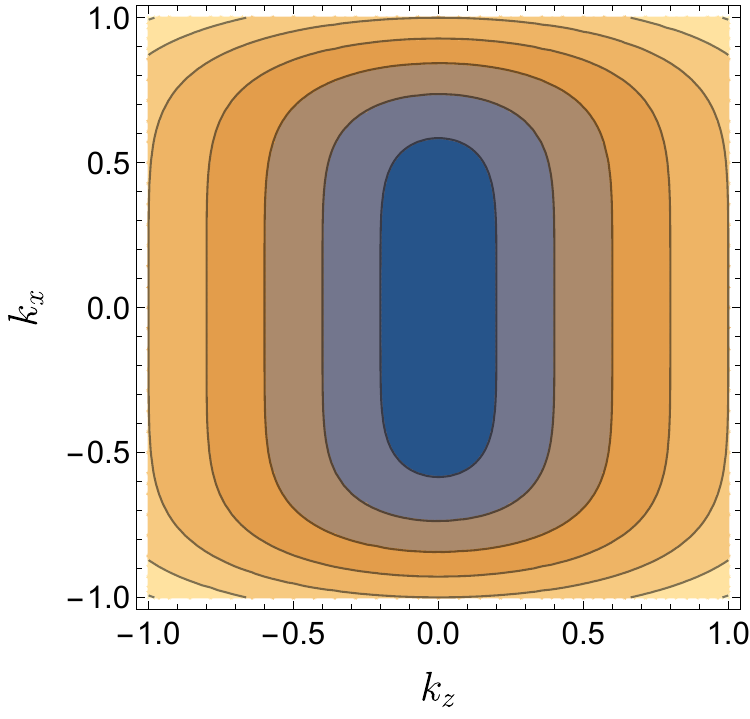}} \end{minipage}& \begin{minipage}{0.5\textwidth} \subfloat[][Slice at $k_x=0$]{\includegraphics[width=0.6\textwidth]{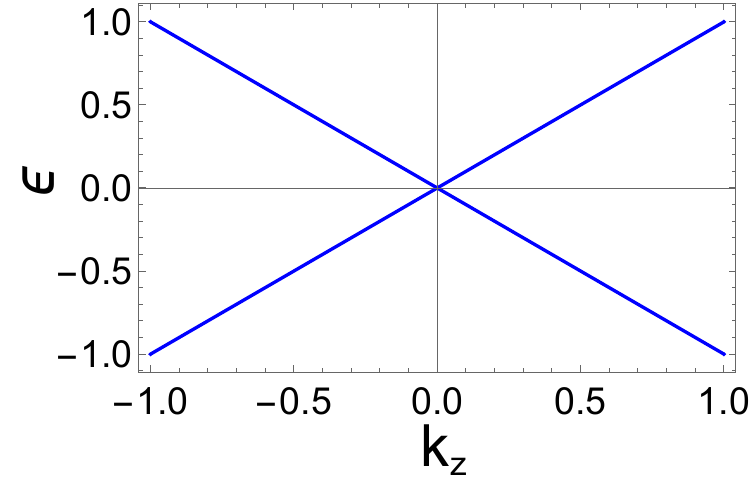}} \\ \subfloat[][Slice at $k_z=0$]{\includegraphics[width=0.6\textwidth]{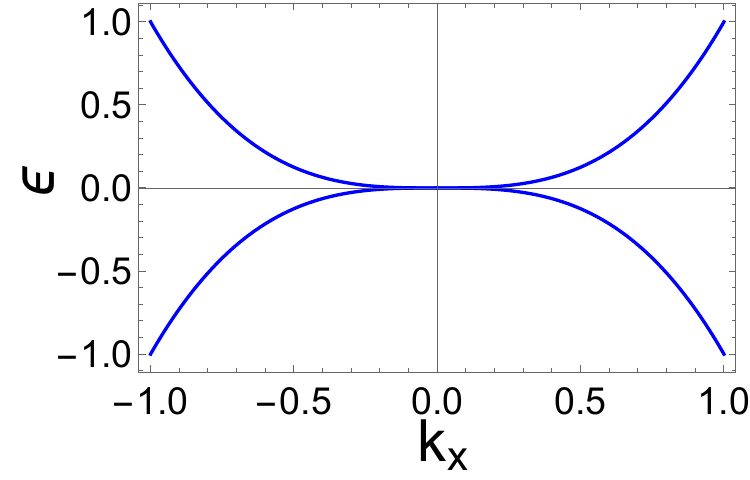}} \end{minipage}
    \end{tabular}
        \caption{\sf (color online) Energy dispersion in the region $z<-L/2$ for $\nu=1$. }
        \label{fig:energy_countour_kz}
    \end{figure}

In this section we consider a bulk cubic Dirac semi-metal and put a potential barrier along the $z$-direction rather than in the plane of the cubic dispersion. That is we consider the Hamiltonian
\begin{equation}
    H=\begin{pmatrix}
    \nu \hat k_z&k_x^3&0&0\\
    k_x^3&-\nu\hat k_z&0&0\\
    0&0&-\nu\hat k_z&-k_x^3\\
    0&0&-k_x^3&\nu\hat k_z
    \end{pmatrix}+u(z)\mathbb{1}_4,
    \label{eq:ham_kz_dir_last_sec}
\end{equation}
where $u(z)=\theta(z+L/2)(1-\theta(z-L/2))$ ($\theta$ is the Heaviside function) and for simplicity we considered only incident momenta $k_x$ and $k_z$. For general combinations of $k_x$, $k_y$ and anisotropy parameter $\lambda$ one may apply a rotation to bring the Hamiltonian to this form. Therefore nothing insightful is lost from the discussion using this simplification (see the appendix \ref{app:rotation} for details).
One finds that the energies in the different regions are given by
\begin{equation}
\epsilon_i=\sqrt{\nu k_z^2 + k_x^6}+u_i,
\end{equation}
which for the region $z<-L/2$ is plotted in Fig. \ref{fig:energy_countour_kz} for visualization.

\begin{figure}[H]
\centering
\includegraphics[width=0.6\linewidth]{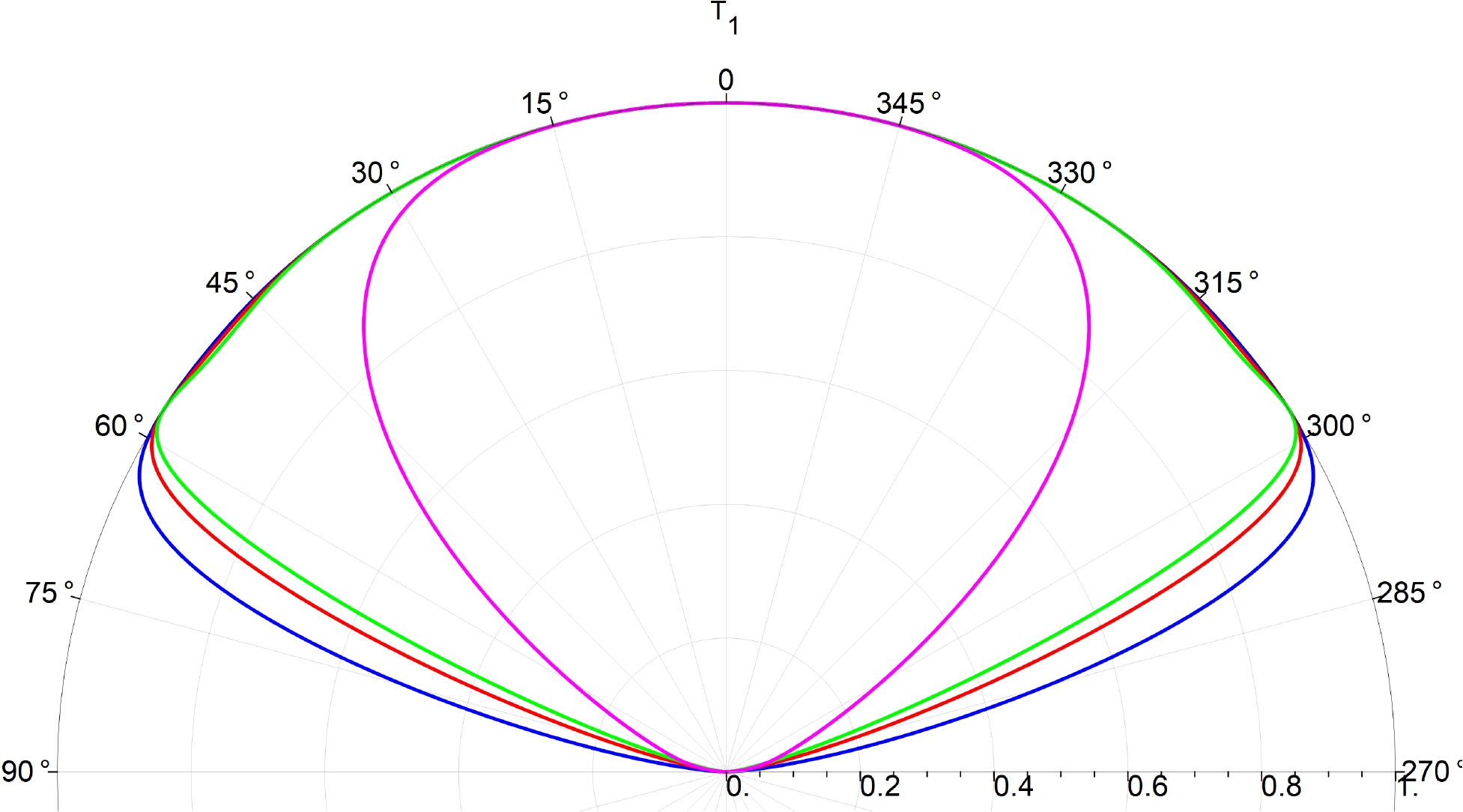}
\caption{\sf (color online) The transmission coefficient $T_1$ for an incident wavevector amplitude $k=1$ and the barrier height $u_0=4$ as a
function  of the incident angle $\theta$ for various values of barrier thickness $L=1$ (blue line), $L=2$ (red line), $L=3$ (green line),$L=3.5$ (magnetia line). 
} \label{fig:fig_transz_angle}
\end{figure}

We find that similarly to the case of graphene \cite{Katsnelson_2006} the scattering problem can be solved by only matching wavefunctions since the barrier is in the linearly dispersive direction. As a result one finds the following transmission amplitudes
\begin{equation}
T_{1,2}=\frac{1}{\cos ^2(L q_{z,i})+\frac{\sin ^2(L q_{z,i}) \left(u^2-\nu ^2 \left(k_{z,i}^2+q_{z,i}^2\right)\right)^2}{4 k_{z,i}^2 \nu ^4 q_{z,i}^2}},
\end{equation}
where we used the definitions $k_x=k\sin(\theta)$, $k_{z,i}=k\cos(\theta)$, $\epsilon=\sqrt{\nu k_z^2+k_x^6}$ and $q_{z,i}=\frac{1}{\nu}\sqrt{(\epsilon-u)^2-k_x^6}$. 
This result may be expanded for small $\theta$ to gain further insight. One then finds that
\begin{equation}
    T_{1,2}\approx1-\frac{  k^4 u^2 \sin ^2\left(\frac{L}{\nu} |u-k\nu |\right)}{\nu ^2 (u-k \nu )^2}\theta^6.
\end{equation}
This result tells us that there is Klein tunneling at small angles and that it is stronger than in the case of graphene and Weyl semi-metals where $T-1\propto \theta^2$. That is for the cubic Dirac semi-metal the Klein tunneling happens over a larger range of angles. A plot of the transmission coefficient as a function of the incident angle $\theta$ can be found in Fig. \ref{fig:fig_transz_angle}.
We find observe that  $T_1$ is very close to unity in a larger angle range $-60^\circ<\theta<60^\circ$.

One should note that this result can be generalized to arbitrary order Dirac semi-metals as follows 
\begin{equation}
    T_{1,2}\approx 1-\frac{ \left[u^2 k^{2 (n-1)} \sin ^2(L (u-k \nu ))\right]}{\nu ^2 (u-k \nu )^2}\theta ^{2 n}.
\end{equation}
Therefore Dirac semi-metals of increasing dispersion order (n) have increasingly strong Klein tunneling.
\section{Conclusion}
\label{sec:conclusion}

In this paper we have investigated the tunneling properties of electrons through a potential barrier in semi-metals described by a cubically dispersed Dirac equation. The cubic Dirac equation was recently predicted based on space-group symmetry considerations and later found to be realizable in a real solid state system, the quasi-one-dimensional transition-metal molybdenum monochalcogenide compound, However, our present work represents the first attempt, to our knowledge, that investigates the theoretical consequences of such a cubically dispersed Dirac system.

Since the considered cubic Dirac equation has many interesting properties such as a linear dispersion in the $z$-direction, cubic dispersion  and rotational symmetry or lack of it in the $x$-$y$ plane, we have designed different situations where each of these effects is isolated and investigated separately. Finally have combined all effects together at a later stage to see if they conspire to create a physical effect that does not occur in isolated situations.

First, we considered the very thin layer limit where the linear dispersion along the $z$-direction is ignored but we maintained the rotational symmetry of the energy dispersion in the $x$-$y$ plane. In this case, we found that the system behaves similarly to single layer graphene or linearly dispersive Dirac semi-metals. In particular, we found that Klein tunneling at normal incidence persists and, much like in these materials, for small angles behaves as $T-1\propto \theta^2$. 
However, at oblique incidence on the square barrier we observed the appearance of Fabry-P\'erot resonances.

Second, we considered the situation with a non-zero momentum eigenvalue $k_z$ which then acts as an effective mass term in the Hamiltonian. In this situation, as expected, we observed a suppression of Klein tunneling except at specific resonant values. 

Third, we considered both previous cases but with the inclusion of an anisotropy parameter $\lambda$ that breaks rotational symmetry. In this regime we found that for different incident angles $\theta$ and anistotropy parameter $\lambda$ we were able to get different numbers of real valued wavevector components  $k_x$. Real valued solutions for $k_x$ are the most important solutions since they represent the propagating modes that can be detected by a distant detector. This allowed us to draw a phase diagram (Fig. \ref{fig:anisotrop_lambda_theta_phase_diag}) that exposes clearly how many modes can be detected by a distant detector as we vary the incidence angle and anisotropy parameter.

Finally, we considered a potential barrier in the linearly dispersive $z$-direction and observed that Klein tunneling occurs over a larger range of angles as compared to graphene, that is for small angles $T-1\propto\theta^6$.

In summary we  demonstrated that cubic Dirac semi-metals host a wealth of interesting tunneling phenomena and that our present theoretical investigation might represent just a tip of an iceberg. We hope, however, that our work paves the way for additional research into transport properties of this phenomenologically rich exotic semi-metal.

\section{Acknowledgements}
We acknowledge the support of King Fahd University of Petroelum and Minerals under Research group project "Transport and semi-classical properties of semi-metals", RGxxxx.
\bibliographystyle{elsarticle-num}
\bibliography{literature}

\appendix
\section{The 2D limit - freezing out \texorpdfstring{$k_z$}{kz}}
\label{app:freeze_out_kz}
In this short appendix we will study for completeness why the momentum in $z$-direction can be expected to freeze out at $k_z=0$ if the sample is sufficiently thin.
Since $v_z$ is the dominant energy scale as a first approximation one can neglect the terms $\propto v_{x,y}$. Therefore we may consider the Hamiltonian
\begin{equation}
    H=\hbar v_z\tau_z\otimes\sigma_z \hat k_z,
\end{equation}
where $\sigma$ are Pauli matrices acting in orbital space and $\tau$ acts in spin space.

For a thin sample we expect electrons to stay localized in the sample and not leak out into the surroundings. Therefore we need a proper description of this and we shall choose the simplest description available for the Hamiltonian. One may recognize quite easily that a potential $V(z)\mathbb{1}_4$ for this type of Hamiltonian cannot localize  particles in $z$-direction inside the sample. The simplest way to localize particles for this type of Hamiltonian is to add a mass-like term of form $M(z)\sigma_{x,y}\otimes \tau_{z,0}$ or similar to the Hamiltonian \cite{Wurm_2011,Akhmerov_2008,doi:10.1098/rspa.1987.0080}. Since this does not couple the different blocks of the Hamiltonian which are copies of each other up to a sign then it is sufficient to consider the different blocks separately. Therefore to understand how the electrons in 
our systems behave when trapped along the $z$-direction it is sufficient to consider the simpler problem of a Hamiltonian
\begin{equation}
     H=\hbar v_z\sigma_z \hat k_z+M(z)\sigma_{x,y},
\end{equation}
where
\begin{equation}
    M(z)=\begin{cases}
    \infty &z<-a/2\\
    0&-a/2<z<a/2\\
    \infty &a/2<z
    \end{cases}.
\end{equation}
For $-a/2<z<a/2$ the Hamiltonian has eigenvectors $V_1=(1,0)e^{-ikz}$ and $V_2=(0,1)e^{ikz}$ that correspond to the same energy $E=\hbar v_z k$. Therefore a general solution at this energy is of the form $\psi=AV_1+BV_2$. It is easy to see \cite{Wurm_2011,doi:10.1098/rspa.1987.0080,Akhmerov_2008} that  possible boundary conditions at $z=\pm a/2$ are 
\begin{equation}
    \psi(\pm a/2)=\sigma_{x,y} \psi(\pm a/2).
\end{equation}
From here one can directly read off that $k$ has to fulfil
\begin{equation}
 e^{2i a k}=1.
\end{equation}
Therefore energy is quantized as
\begin{equation}
    E=\hbar v_z\pi \frac{n}{a}; \quad n\in \mathbb{Z}.
\end{equation}
Consequently, for sufficiently small $a$ the distance between energies is very large. Therefore at low energies $\left|E\right|<\hbar v_z\pi/a$ one may set $k_z=0$ in the main text.
\section{Wavefunctions in the different spatial regions}
\subsection{Wavefunctions for the isotropic \texorpdfstring{($k_z\neq 0$)}{(kz=0) }  limit}
\label{app:wavefunctions_mass_term}
To find the wavefunctons in the case of the massive rotation symmetric case we may now repeat the analysis from Sec. \ref{sec:Sol:wf:massless_isotrol} for which some of the $k_{x,n}^\pm$ are physically allowed to find an ansatz for the wavefunction in \textbf{region 1}. The ansatz is given by
\begin{equation}
\begin{aligned}
&\Phi_{1}(x)=a_1 e_{k,1}^+\begin{pmatrix}
\tilde \chi_{k,1}^{+(+)} \\ 1 \\ 0 \\ 0
\end{pmatrix}+ a_2 e_{k,1}^+\begin{pmatrix}
0 \\ 0 \\ \tilde \chi_{k,1}^{+(-)} \\ 1
\end{pmatrix} +a_3 e_{k,2}^+ \begin{pmatrix}
\tilde \chi_{k,2}^{+(+)} \\ 1 \\ 0 \\ 0
\end{pmatrix}+ a_4 e_{k,2}^+ \begin{pmatrix}
0 \\ 0 \\ \tilde \chi_{k,2}^{+(-)} \\ 1
\end{pmatrix} + a_5 e_{k,1}^-\begin{pmatrix}
\tilde \chi_{k,1}^{-(+)} \\ 1 \\ 0 \\ 0
\end{pmatrix} \\
& + a_6 e_{k,1}^-\begin{pmatrix}
0 \\ 0 \\ \tilde \chi_{k,1}^{-(-)} \\ 1
\end{pmatrix}  + a_7  e_{k,3}^-\begin{pmatrix}
\tilde \chi_{k,3}^{-(+)} \\ 1 \\ 0 \\ 0
\end{pmatrix}     + a_8 e_{k,3}^-\begin{pmatrix}
0 \\ 0 \\\tilde \chi_{k,3}^{-(-)}, \\ 1
\end{pmatrix},
\label{eq:phi1delta}
\end{aligned}
\end{equation}
where we used the shorthand notation $e_{k,n}^\pm=e^{i k_{x,n}^\pm x}$ and defined 
\begin{equation}
    \tilde \chi_{k,n}^{\pm(\pm)}=\frac{\delta+(\pm) \sqrt{({k_{x,n}^\pm}^2+(k \sin\theta)^2)^3+ \delta^2}}{(k_{x,n}^\pm - i k \sin\theta)^{3}}.
    \label{eq:shorthandchitilde}
    \end{equation}

In \textbf{region 2}  we have a modified wavevector amplitude $q=\left[(\epsilon-u)^2-\delta^2\right]^{1/6}$  and angle $\phi=\sin^{-1}\left(\frac{k}{q}\sin\theta\right)$, which are found by enforcing energy and momentum conservation in $y$-direction. The $q_x$ component eigenvalues are like in the previous section given by Eq. \eqref{eq:k1} with $k\to q$ and $\theta\to \phi$. In this region, all of the $k_n^\pm$ are physically allowed, which dictates that the ansatz wavefunction in this region is given by
\begin{equation}
\begin{aligned}
 \label{eq:phi2-delta0-anisotrop}
&\Phi_{2}(x)= b_1 e_{q,1}^+\begin{pmatrix}
\tilde\chi_{q,1}^{+(-)} \\ 1 \\ 0 \\ 0
\end{pmatrix}+b_2 e_{q,1}^+\begin{pmatrix}
0 \\ 0 \\ \tilde \chi_{q,1}^{+(+)} \\ 1
\end{pmatrix} +  b_3 e_{q,2}^+\begin{pmatrix}
\tilde \chi_{q,2}^{+(-)} \\ 1 \\ 0 \\ 0
\end{pmatrix}+  b_4 e_{q,2}^+\begin{pmatrix}
0 \\ 0 \\ \tilde \chi_{q,2}^{+(+)} \\ 1
\end{pmatrix}+  b_5 e_{q,3}^+\begin{pmatrix}
\tilde \chi_{q,3}^{+(-)} \\ 1 \\ 0 \\ 0
\end{pmatrix}+  b_6 e_{q,3}^+\begin{pmatrix}
0 \\ 0 \\ \tilde \chi_{q,3}^{+(+)} \\ 1
\end{pmatrix}\\
&+  b_7 e_{q,1}^-\begin{pmatrix}
 \tilde \chi_{q,1}^{-(-)} \\ 1 \\ 0 \\ 0
\end{pmatrix}+  b_8 e_{q,1}^-\begin{pmatrix}
0 \\ 0 \\ \tilde \chi_{q,1}^{-(+)} \\ 1
\end{pmatrix} +  b_9 e_{q,2}^-\begin{pmatrix}
\tilde \chi_{q,2}^{-(-)} \\ 1 \\ 0 \\ 0
\end{pmatrix}+  b_{10} e_{q,2}^-\begin{pmatrix}
0 \\ 0 \\ \tilde \chi_{q,2}^{-(+)} \\ 1
\end{pmatrix} +  b_{11} e_{q,3}^-\begin{pmatrix}
\tilde \chi_{q,3}^{-(-)} \\ 1 \\ 0 \\ 0
\end{pmatrix}  +  b_{12} e_{q,3}^-\begin{pmatrix}
0 \\ 0 \\ \tilde \chi_{q,3}^{-(+)} \\ 1
\end{pmatrix}
\end{aligned},
\end{equation}
where we made use of the shorthand notation \eqref{eq:shorthandchitilde}.

For \textbf{region 3} $\left(x>\frac{L}{2}\right)$, where only propagation away from the barrier is allowed and terms need to be finite for $x\to + \infty$, we find that only $k_{x,1}^+$, $k_{x,2}^-$ and $k_{x,3}^+$ fulfil these properties. Therefore, an ansatz for the spinor in this region is given by 
\begin{equation}
\begin{aligned}
&\Phi_{3}(x) = t_1 e_{k,1}^+ \begin{pmatrix}
\tilde \chi_{k,1}^{+(+)} \\ 1 \\ 0 \\ 0
\end{pmatrix}+t_2 e_{k,1}^+\begin{pmatrix}
0 \\ 0 \\ \tilde \chi_{k,1}^{+(-)} \\ 1
\end{pmatrix} + t_3 e_{k,3}^+ \begin{pmatrix}
\tilde \chi_{k,3}^{+(+)} \\ 1 \\ 0 \\ 0
\end{pmatrix}+  t_4 e_{k,3}^+ \begin{pmatrix}
0 \\ 0 \\ \tilde \chi_{k,3}^{+(-)}\\ 1
\end{pmatrix}+  t_5  e_{k,2}^- \begin{pmatrix}
\tilde \chi_{k,2}^{-(+)} \\ 1 \\ 0 \\ 0
\end{pmatrix}
 +  t_6  e_{k,2}^- \begin{pmatrix}
0 \\ 0 \\ \tilde \chi_{k,2}^{-(-)} \\ 1
\end{pmatrix}.
\label{eq:phi3-delta0-anisotrop} 
\end{aligned}
\end{equation}
 \subsection{Wavefunctions for the anisotropic \texorpdfstring{$(k_z=0)$}{(kz=0)} limit}
 \label{app:Wavefunct_non_rot_symm_zero_mass}
 
Much like previously also in the case of broken rotational symmetry we have to choose wavefunctions that correspond to wavevectors that  are physically allowed. By doing so we find that the solution in \textbf{region 1} is  given as
\begin{equation}
\begin{aligned}
&\Phi_{1}(x)= a_1 e_{k,1}^+\begin{pmatrix}
\chi_{k,1}^{+} \\ 1 \\ 0 \\ 0
\end{pmatrix} +a_2 e_{k,1}^+\begin{pmatrix}
0 \\ 0 \\ - \chi_{k,1}^{+} \\ 1
\end{pmatrix} + a_3 e_{k,2}^+\begin{pmatrix}
\chi_{k,2}^{+} \\ 1 \\ 0 \\ 0
\end{pmatrix}+  a_4 e_{k,2}^+ \begin{pmatrix}
0 \\ 0 \\ - \chi_{k,2}^{+} \\ 1
\end{pmatrix} 
+a_5 e_{k,1}^-\begin{pmatrix}
\chi_{k,1}^{-} \\ 1 \\ 0 \\ 0
\end{pmatrix} \\
&+ a_6 e_{k,1}^-\begin{pmatrix}
0 \\ 0 \\ -\chi_{k,1}^{-} \\ 1
\end{pmatrix}  + a_7  e_{k,3}^-\begin{pmatrix}
 \chi_{k,3}^{-} \\ 1 \\ 0 \\ 0
\end{pmatrix} + a_8 e_{k,3}^-\begin{pmatrix}
0 \\ 0 \\ -\chi_{k,3}^{-} \\ 1
\end{pmatrix},
\end{aligned}
 \label{eq:phi1}
\end{equation}
where we have used the shorthand notation
\begin{equation}
e_{k,n}^\pm=e^{ i k_{x,n}^\pm x},
\label{eq:anisotropdelta0expdef}
\end{equation} 
and
\begin{equation}
\chi_{k,n}^{\pm}=\frac{\sqrt{\lambda^2 (k_{x,n}^\pm)^2 \left((k_{x,n}^\pm)^2-3 k_y^2\right)^2+\left(k_y^3-3 (k_{x,n}^\pm)^2 k_y\right)^2}}{\lambda (k_{x,n}^\pm)^3-3 i (k_{x,n}^\pm)^2 k_y-3 \lambda (k_{x,n}^\pm) k_y^2+i k_y^3}.
\label{eq:anisotropdelta0chidef}
\end{equation} 
with $k_y=k\sin\theta$.

Also in \textbf{region 2} $\left(-\frac{L}{2}<x<\frac{L}{2}\right)$  one may use Eq. \eqref{eq:anisotropkz0kxeigeneq} to determine the allowed  $q_x$ (renaming $k_x\to q_x$) for a given value of $k_y=k\sin\theta$.  We may then use one of the $q_x$ (see
more on this below) to define the wavevector amplitude $q=\sqrt{q_x^2+k_y^2}$ and angle $\phi=\arcsin\left(\frac{k\sin\theta}{q}\right)$. If we choose the $q_x$ solution (proper care is needed) one may then use Eq. \eqref{eq:anisotrop_k1} with the replacements $k\to q$ and $\theta\to \phi$ to have all the $q_x$ solutions expressed in a compact form. 
The wavefunction in this region  
is then given by
\begin{equation}  
\begin{aligned}
&\Phi_{2}(x)= b_1 e_{q,1}^+\begin{pmatrix}
-\chi_{q,1}^{+} \\ 1 \\ 0 \\ 0
\end{pmatrix} + b_2 e_{q,1}^+\begin{pmatrix}
0 \\ 0 \\ \chi_{q,1}^{+} \\ 1
\end{pmatrix} +  b_3 e_{q,2}^+\begin{pmatrix}
-\chi_{q,2}^{+} \\ 1 \\ 0 \\ 0
\end{pmatrix} +  b_4 e_{q,2}^+\begin{pmatrix}
0 \\ 0 \\ \chi_{q,2}^{+} \\ 1
\end{pmatrix}+  b_5 e_{q,3}^+\begin{pmatrix}
-\chi_{q,3}^{+} \\ 1 \\ 0 \\ 0
\end{pmatrix} + b_6 e_{q,3}^+\begin{pmatrix}
0 \\ 0 \\ \chi_{q,3}^{+} \\ 1
\end{pmatrix} \\
&+  b_7 e_{q,1}^-\begin{pmatrix}
-\chi_{q,1}^{-} \\ 1 \\ 0 \\ 0
\end{pmatrix}+  b_8 e_{q,1}^-\begin{pmatrix}
0 \\ 0 \\ \chi_{q,1}^{-} \\ 1
\end{pmatrix} 
+  b_9 e_{q,2}^-\begin{pmatrix}
-\chi_{q,2}^{-} \\ 1 \\ 0 \\ 0
\end{pmatrix}+  b_{10} e_{q,2}^-\begin{pmatrix}
0 \\ 0 \\ \chi_{q,2}^{-} \\ 1
\end{pmatrix} +  b_{11} e_{q,3}^-\begin{pmatrix}
-\chi_{q,3}^{-} \\ 1 \\ 0 \\ 0
\end{pmatrix} +   +  b_{12} e_{q,3}^-\begin{pmatrix}
0 \\ 0 \\ \chi_{q,3}^{-} \\ 1
\end{pmatrix},
\end{aligned}
\label{eq:phi2}
\end{equation}
where we used the same shorthand notations as in {\bf region 1},  Eqs. (\ref{eq:anisotropdelta0expdef}-\ref{eq:anisotropdelta0chidef}), just with replacements $k\to q$ and $\theta\to\phi$.

For \textbf{region 3} $\left(x>\frac{L}{2}\right)$ we may use all definitions from {\bf region 1} without alteration and find a wavefunction ansatz as
\begin{equation}
\begin{aligned}
\label{eq:phi3} 
&\Phi_{3}(x) =t_1 e_{k,1}^+ \begin{pmatrix}
\chi_{k,1}^{+} \\ 1 \\ 0 \\ 0
\end{pmatrix} + t_2 e_{k,1}^+\begin{pmatrix}
0 \\ 0 \\ -\chi_{k,1}^{+} \\ 1
\end{pmatrix}+ t_3 e_{k,3}^+ \begin{pmatrix}
\chi_{k,3}^{+} \\ 1 \\ 0 \\ 0
\end{pmatrix}+  t_4 e_{k,3}^+ \begin{pmatrix}
0 \\ 0 \\ -\chi_{k,3}^{+}\\ 1
\end{pmatrix} + t_5  e_{k,2}^- \begin{pmatrix}
\chi_{k,2}^{-} \\ 1 \\ 0 \\ 0
\end{pmatrix} +  t_6  e_{k,2}^- \begin{pmatrix}
0 \\ 0 \\ -\chi_{k,2}^{-} \\ 1
\end{pmatrix}.
\end{aligned}
\end{equation}
\subsection{Wavefunctions for the anisotropic \texorpdfstring{$(k_z\neq0)$}{(kz=0)} limit}
\label{app:Wavefunct_non_rot_symm_massive}
 For \textbf{region 1} $\left(x<-\frac{L}{2}\right)$, we find that only some of the $k_{x,i}$ and their corresponding eigenvectors are allowed by the boundary conditions. That is the wavefunctions have to be finite as $x\to-\infty$. We find that the most general such wavefunction is given by 
\begin{equation}
\begin{aligned}
&\Phi_{1}(x)= 
a_1 e_{k,1}^+\begin{pmatrix}
\chi_{k,1}^{+(+)} \\ 1 \\ 0 \\ 0
\end{pmatrix} +a_2 e_{k,1}^+\begin{pmatrix}
0 \\ 0 \\ \chi_{k,1}^{-(+)} \\ 1
\end{pmatrix}+ a_3 e_{k,2}^+\begin{pmatrix}
\chi_{k,2}^{+(+)} \\ 1 \\ 0 \\ 0
\end{pmatrix}
+  a_4 e_{k,2}^+ \begin{pmatrix}
0 \\ 0 \\ \chi_{k,2}^{+(+)} \\ 1
\end{pmatrix} 
+a_5 e_{k,1}^-\begin{pmatrix}
\chi_{k,1}^{+(-)} \\ 1 \\ 0 \\ 0
\end{pmatrix}\\
&+ a_6 e_{k,1}^-\begin{pmatrix}
0 \\ 0 \\ \chi_{k,1}^{-(-)} \\ 1
\end{pmatrix}
+ a_7  e_{k,3}^-\begin{pmatrix}
 \chi_{k,3}^{+(-)} \\ 1 \\ 0 \\ 0
\end{pmatrix} + a_8 e_{k,3}^-\begin{pmatrix}
0 \\ 0 \\ \chi_{k,3}^{-(-)} \\ 1
\end{pmatrix},
\end{aligned}
 \label{eq:gappedanisotrop_phi1}
\end{equation}
where we have used the shorthand notation
\begin{equation}
e_{k,n}^\pm=e^{ i k_{x,n}^\pm x},
\end{equation} 
and 
\begin{equation}
\begin{aligned}
&\chi_{k,n}^{l\pm}=\frac{\delta+l \sqrt{\lambda^2 (k_{x,n}^\pm)^2 \left[(k_{x,n}^\pm)^2-3 k_y^2\right]^2+\left[k_y^3-3 (k_{x,n}^\pm)^2 k_y\right]^2+\delta^2}}{\lambda (k_{x,n}^\pm)^3-3 i (k_{x,n}^\pm)^2 k_y-3 \lambda (k_{x,n}^\pm) k_y^2+i k_y^3}
\end{aligned}
\end{equation} 
where $l=\pm1$ and $k_y=k\sin\theta$.

In \textbf{region 2}  $\left(-\frac{L}{2}<x<\frac{L}{2}\right)$  we also have to determine possible wavevectors by solving Eq. \eqref{eq:anisotropkznot0kxeigeneq} for $q_x$ (renaming $k_x\to q_x$). These solutions can also be written in the compact form \eqref{eq:anisotrop_k1} if we replace $k\to q$ and $\theta\to \phi$. Here we have to define $\phi=\arcsin\left(\frac{k\sin\theta}{q}\right)$ and $q=\sqrt{q_x^2+k_y^2}$. Of course proper care has to be taken to choose the proper real valued solution of $q_x$ that allows us to properly reproduce all solutions to \eqref{eq:anisotropkznot0kxeigeneq}. With these expressions in mind an ansatz for this region  can be chosen to have the form
\begin{equation} 
\begin{aligned}
&\Phi_{2}(x)=b_1 e_{q,1}^+\begin{pmatrix}
\chi_{q,1}^{-(+)} \\ 1 \\ 0 \\ 0
\end{pmatrix} + b_2 e_{q,1}^+\begin{pmatrix}
0 \\ 0 \\ \chi_{q,1}^{+(+)} \\ 1
\end{pmatrix} +  b_3 e_{q,2}^+\begin{pmatrix}
\chi_{q,2}^{-(+)} \\ 1 \\ 0 \\ 0
\end{pmatrix} +  b_4 e_{q,2}^+\begin{pmatrix}
0 \\ 0 \\ \chi_{q,2}^{+(+)} \\ 1
\end{pmatrix}+  b_5 e_{q,3}^+\begin{pmatrix}
\chi_{q,3}^{-+} \\ 1 \\ 0 \\ 0
\end{pmatrix} + b_6 e_{q,3}^+\begin{pmatrix}
0 \\ 0 \\ \chi_{q,3}^{+(+)} \\ 1
\end{pmatrix}\\
&+  b_7 e_{q,1}^-\begin{pmatrix}
\chi_{q,1}^{-(-)} \\ 1 \\ 0 \\ 0
\end{pmatrix}+  b_8 e_{q,1}^-\begin{pmatrix}
0 \\ 0 \\ \chi_{q,1}^{+(-)} \\ 1
\end{pmatrix} 
+  b_9 e_{q,2}^-\begin{pmatrix}
\chi_{q,2}^{-(-)} \\ 1 \\ 0 \\ 0
\end{pmatrix}+  b_{10} e_{q,2}^-\begin{pmatrix}
0 \\ 0 \\ \chi_{q,2}^{+(-)} \\ 1
\end{pmatrix} +  b_{11} e_{q,3}^-\begin{pmatrix}
\chi_{q,3}^{--} \\ 1 \\ 0 \\ 0
\end{pmatrix} +   +  b_{12} e_{q,3}^-\begin{pmatrix}
0 \\ 0 \\ \chi_{q,3}^{+(-)} \\ 1
\end{pmatrix}
\end{aligned}.
\label{eq:gappedanisotrop_phi2}
\end{equation}

Last, in \textbf{region 3} $\left( x>\frac{L}{2}\right)$, taking into account that the wavefunction needs to be finite as $x\to +\infty$, we may therefore consider the following ansatz
\begin{equation}
    \begin{aligned}
&\Phi_{3}(x) =t_1 e_{k,1}^+ \begin{pmatrix}
\chi_{k,1}^{+(+)} \\ 1 \\ 0 \\ 0
\end{pmatrix} + t_2 e_{k,1}^+\begin{pmatrix}
0 \\ 0 \\ \chi_{k,1}^{-(+)} \\ 1
\end{pmatrix}+ t_3 e_{k,3}^+ \begin{pmatrix}
\chi_{k,3}^{+(+)} \\ 1 \\ 0 \\ 0
\end{pmatrix}+  t_4 e_{k,3}^+ \begin{pmatrix}
0 \\ 0 \\ \chi_{k,3}^{-(+)}\\ 1
\end{pmatrix} + t_5  e_{k,2}^- \begin{pmatrix}
\chi_{k,2}^{+(-)} \\ 1 \\ 0 \\ 0
\end{pmatrix} +  t_6  e_{k,2}^- \begin{pmatrix}
0 \\ 0 \\ \chi_{k,2}^{-(-)} \\ 1
\end{pmatrix}
\end{aligned}
\label{eq:gappedanisotrop_phi3} 
\end{equation}
\section{Rotation for the tunneling Hamiltonian in z-direction}
\label{app:rotation}

In this appendix we discuss how the Hamiltonian can for the tunneling in the $z$-direction be brought into the form of equation \eqref{eq:ham_kz_dir_last_sec}.
The Hamiltonian that one may consider for the more general case of a wave that has vector components $(k_x,k_y)=k\sin\theta(\cos\psi,\sin\psi)$ with an angle $\psi$ in the plane of the barrier is given as
 \begin{equation}
  H(k) = \left(\begin{array}{cccc} v_z/v k_z& \hat{k}_{1}&0&0\\ \hat{k}_{1}^\dagger&0&0&0\\0&0&0&- \hat{k}_{1}\\0&0&-\hat{k}_{1}^\dagger&0\\\end{array}\right)+u_j\mathbb{1}_4,
\end{equation}
where $k_1=\frac{\lambda+1}{2}k_+^3+\frac{\lambda-1}{2}k_-^3$ as before.
Through a unitary transformation $U^\dag H(k)U$ one can bring the Hamiltonian into a form that only has momenta along the $z$ and $x$ directions. The transformation is given as
\begin{equation}
    U=\mathbb{1}_2\otimes e^{i/2\sigma_3\gamma};\quad \gamma=\arctan(\lambda^{-1}\tan(3\psi)),
\end{equation}
where we used the usual notation adopted in the main text. The transformation corresponds to a rotation with angle $\gamma$ around the $z$-direction. With these changes we find equation \eqref{eq:ham_kz_dir_last_sec} if we define the constant
\begin{equation}
    \nu=\frac{v_z}{v}\left(\lambda  \cos (3 \psi ) \sqrt{\frac{\tan ^2(3 \psi )}{\lambda ^2}+1}\right)^{-1}
\end{equation}
and make use of a reduced energy that is defined slightly differently than in the rest of the text as indicated below
\begin{equation}
    \epsilon=\frac{E}{2\hbar v}\left(\lambda  \cos (3 \psi ) \sqrt{\frac{\tan ^2(3 \psi )}{\lambda ^2}+1}\right)^{-1}.
\end{equation}
 
\end{document}